\documentclass[12pt]{JHEP3}
\usepackage{epsfig,latexsym,amsmath,amssymb,bm,cite}
\usepackage{graphicx}

\newcommand{\be}{\begin{equation}}
\newcommand{\ee}{\end{equation}}
\newcommand{\bea}{\begin{eqnarray}}
\newcommand{\eea}{\end{eqnarray}}
\newcommand{\bem}{\begin{multline}}
\newcommand{\eem}{\end{multline}}
\newcommand{\beg}{\begin{gather}}
\newcommand{\eeg}{\end{gather}}

\def\eq#1{{Eq.~(\ref{#1})}}
\def\fig#1{{Fig.~\ref{#1}}}
\newcommand{\ben}{\begin{eqnarray*}}
\newcommand{\een}{\end{eqnarray*}}

\title{Asymmetric Collision of Two Shock Waves in AdS$_{\bf 5}$}

\author{ Javier L.\ Albacete${}^{\,1}$, \ Yuri V.\ Kovchegov${}^{\,2}$, \ 
Anastasios Taliotis${}^{\,2}$ \\
\vspace{0.1in}

${}^{\,1}$ECT*, Strada delle Tabarelle 286, I-38050, Villazzano (TN), Italy 
\vspace{0.1in}

${}^{\,2}$Department of Physics, The Ohio State University,
Columbus, OH 43210, USA 
\vspace{0.1in}

E-mail addresses: \email{albacete@mps.ohio-state.edu}, 
\email{yuri@mps.ohio-state.edu}, \email{taliotis.1@osu.edu}
}

\date{February 2009}

\abstract{We consider high energy collisions of two shock waves in
  AdS$_5$ as a model of ultrarelativistic nucleus-nucleus collisions
  in the boundary theory. We first calculate the graviton field
  produced in the collisions in the NLO and NNLO approximations,
  corresponding to three- and four-graviton exchanges with the shock
  waves. We then consider the asymmetric limit where the energy
  density in one shock wave is much higher than in the other one. In
  the boundary theory this setup corresponds to proton--nucleus
  collisions, with the nucleus being the denser of the two shock waves
  and the proton being the less dense one. Employing the eikonal
  approximation we find the exact high energy analytic solution for
  the metric in AdS$_5$ for the asymmetric collision of two
  delta-function shock waves. The solution resums all-order graviton
  exchanges with the ``nucleus'' shock wave and a single-graviton
  exchange with the ``proton'' shock wave. Using the holographic
  renormalization prescription we read off the energy-momentum tensor
  of the matter produced in proton-nucleus collisions. We show in
  explicit detail that in the boundary theory the proton is completely
  stopped by strong-coupling interactions with the nucleus, in
  agreement with our earlier results \cite{Albacete:2008vs}. We also
  apply the eikonal technique to the asymmetric collision of two
  unphysical delta-prime shock waves, which we introduced in
  \cite{Albacete:2008vs} as a means of modeling nuclear collisions
  with weak coupling initial dynamics. We obtain a surprising result
  that, for delta-prime shock waves, the multiple bulk graviton
  exchange series giving the leading energy-dependent contribution to
  the energy-momentum tensor terminates at the order of two graviton
  exchanges with the nucleus.  }


\preprint{ECT*-09-02}

\begin{document}


\section{Introduction}

In this paper we continue our earlier investigation
\cite{Albacete:2008vs} of colliding shock waves in AdS$_5$. Due to the
Anti-de Sitter space/conformal field theory (AdS/CFT) correspondence
\cite{Maldacena:1997re,Gubser:1998bc,Witten:1998qj,Aharony:1999ti},
the problem of two colliding gravitational shock waves, while an
important problem from the standpoint of gravity theory
\cite{D'Eath:1992hb,D'Eath:1992hd,D'Eath:1992qu,Sperhake:2008ga}, may
also be relevant for high energy hadronic and nuclear collisions at
strong coupling
\cite{Nastase:2005rp,Kajantie:2008rx,Grumiller:2008va,Gubser:2008pc,Lin:2009pn}.

One of the most important problems in the field of ultrarelativistic
heavy ion collisions is the one of isotropization and thermalization
of the produced medium. There is a growing consensus in the heavy ion
community that the medium produced in heavy ion collisions at RHIC is
strongly coupled
\cite{Kolb:2000sd,Kolb:2000fh,Huovinen:2001cy,Kolb:2001qz,Heinz:2001xi,Teaney:1999gr,Teaney:2000cw,Teaney:2001av,Teaney:2003kp}.
The challenge for the theoretical community is to understand (i) how
the medium, which is initially very anisotropic with zero or negative
longitudinal component of the energy-momentum tensor
\cite{Krasnitz:2003jw,Krasnitz:2002mn,Lappi:2003bi,Fukushima:2007ja,Kovchegov:2007pq,Kovchegov:2005ss},
evolves into an isotropic medium described by ideal (Bjorken
\cite{Bjorken:1982qr}) hydrodynamics, and (ii) why this transition
happens over extremely short time scale of $0.3 \div 0.6$~fm/c, as
required by hydrodynamic simulations
\cite{Kolb:2000sd,Kolb:2000fh,Huovinen:2001cy,Kolb:2001qz,Heinz:2001xi,Teaney:1999gr,Teaney:2000cw,Teaney:2001av}.

There is also a widespread belief in the community, supported by a
broad range of phenomenological evidence, that the very early stages
of heavy ion collisions are weakly-coupled, i.e., they are described
by the physics of Color Glass Condensate (CGC)/parton saturation
\cite{Blaizot:1987nc,McLerran:1993ni,McLerran:1993ka,McLerran:1994vd,Kovchegov:1996ty,Kovchegov:1997pc,Kovner:1995ja,Kovchegov:1997ke,Krasnitz:1998ns,Krasnitz:1999wc,Krasnitz:2003jw,Kovchegov:2000hz,Lappi:2003bi,Kharzeev:2000ph,Kharzeev:2001yq,Kharzeev:2002pc,Kharzeev:2004yx,Albacete:2003iq,Albacete:2007sm}
(for a review of CGC see
\cite{Iancu:2003xm,Weigert:2005us,Jalilian-Marian:2005jf}). It appears
that the system produced in heavy ion collisions evolves with time
from the weakly-coupled CGC state to the strongly coupled quark-gluon
plasma (QGP) described by the ideal hydrodynamics. There are two types
of transitions that the system has to undergo in order for such
process to take place. First of all, at some point in time the system
should undergo a transition from weak coupling to strong coupling.
Second of all, at a (presumably) different time the system will evolve
from the anisotropic early state, in which transverse and longitudinal
pressure components in the energy-momentum tensor are drastically
different, to the isotropic later state, in which all pressure
components are equal (or almost equal) to each other, as required by
the ideal (viscous) hydrodynamics. We will refer to the latter
transition as the {\sl isotropization transition}.  The isotropization
transition is a necessary condition for the thermalization of the
produced medium.

In this work we will assume that the isotropization transition takes
place after the strong coupling transition. Hence we will study the
onset of the isotropization in the strongly-coupled framework. Since
strong coupling dynamics in QCD is prohibitively complicated,
especially for the ultrarelativistic processes at hand, we will employ
AdS/CFT correspondence, assuming that the bulk properties of the
collisions and the produced medium in ${\cal N} =4$ super-Yang-Mills
theory are not too different from QCD and would allow us to make
conclusions which are at least qualitatively applicable to the real
life.

Attempts to study isotropization and thermalization in the AdS/CFT
framework have been made before. A gravity-dual of Bjorken
hydrodynamics was constructed in
\cite{Janik:2005zt,Janik:2006ft,Heller:2007qt,Benincasa:2007tp,Heller:2008mb}.
To obtain it the authors of \cite{Janik:2005zt} assumed that the
medium produced in heavy ion collisions is rapidity-independent.
Imposing a no-singularities requirement \cite{Janik:2005zt} (or simply
demanding that the metric is real \cite{Kovchegov:2007pq}) one then
obtains the asymptotic late-time geometry corresponding to Bjorken
hydrodynamics. However, this result by itself does not prove that
Bjorken hydrodynamics is a consequence of a heavy ion collision. In
other words, it is not clear which early-time dynamics (or, in
general, which events in the past) lead to this dual-Bjorken geometry.

To address this problem, by analogy with the perturbative approaches
\cite{Kovner:1995ts,Kovner:1995ja,Krasnitz:2003jw,Krasnitz:2002mn}, it
was suggested in \cite{Nastase:2005rp} that one should study
collisions of two shock waves in AdS space: following the dynamics of
the strongly-coupled medium produced in such collisions one would be
able to see how the ideal hydrodynamic state is reached by the medium
and whether this late-time state is rapidity-independent. In
\cite{Kajantie:2008rx} the case of shock wave collisions in the $1+1$
dimensional boundary theory was considered and solved exactly in
AdS$_3$ geometry.  Unfortunately the lower dimensionality of the
problem severely limits the physical behavior of the produced medium,
and does not allow to formulate the problem of isotropization. The
case of realistic $1+3$ boundary theory was first addressed in
\cite{Grumiller:2008va} using AdS$_5$ space with the infinitely-thin
delta-function shock waves. The authors of \cite{Grumiller:2008va}
constructed a perturbative series for the energy-momentum tensor of
the produced strongly coupled matter.

In \cite{Albacete:2008vs} we generalized the results of
\cite{Grumiller:2008va} by solving Einstein equations in a more
general framework, which does not depend on the exact profile of the
shock waves, i.e., whether they are delta-functions or some other
objects with finite extent. We identified the perturbation series of
\cite{Grumiller:2008va} with a series in bulk graviton exchanges with
two shock waves (see e.g. \fig{AA} below for an example of a term
contributing to the series). Most importantly, in
\cite{Albacete:2008vs} it was argued that in a collision of any two
physical shock waves, they stop shortly after the collision, possibly
forming a black hole. In the boundary theory this behavior corresponds
to the colliding nuclei stopping shortly after the collision, probably
leading to Landau hydrodynamics description of the system
\cite{Landau:1953gs}. Such complete nuclear stopping would lead to
complete stopping of the baryon number carried by the nuclei.  As such
a complete baryon stopping is not observed at RHIC (and, in fact,
baryon stopping at mid-rapidity at RHIC is rather small
\cite{Bearden:2003hx} in accord with perturbative calculations
\cite{Itakura:2003jp,Albacete:2006vv}), this indicates that colliding
shock waves may not be adequate for the description of realistic
nuclear collisions in AdS. Indeed, an AdS description would apply if
the collisions were strongly-coupled at all times: as the early stages
of RHIC heavy ion collisions are weakly coupled, an AdS/CFT
description of the collision at all times can not be valid.  In an
attempt to resolve the issue we suggested in \cite{Albacete:2008vs}
that one could use unphysical shock waves with the delta-prime
profile. Such shock waves appear to have no stopping.  It is possible
that using delta-prime shock waves as external sources for the AdS/CFT
correspondence would yield a more realistic description of heavy ion
collisions, and would allow one to tackle the problem of
isotropization in the strongly-coupled framework.

In this paper we further explore shock wave collisions. In Sect.
\ref{pert} we extend the expansion in graviton exchanges from
\cite{Albacete:2008vs} to two higher orders. We calculate the
next-to-leading order (NLO) and next-to-next-to-leading order (NNLO)
corrections to the result of \cite{Albacete:2008vs} for both
delta-function and delta-prime shock waves (see Eqs. (\ref{pNLO} and
(\ref{ppNLO}), along with \eq{H2}).

We continue in Sect. \ref{pA} by constructing the resummation
procedure in which graviton exchanges with one shock wave are resummed
to all orders while the interaction with another shock wave is
restricted to a single graviton exchange (see \fig{pAfig} below). The
diagrams are analogous to those resummed in the study of classical
gluon fields produced in proton-nucleus collisions in the perturbative
CGC framework
\cite{Kovchegov:1998bi,Kopeliovich:1998nw,Dumitru:2001ux,Kovchegov:2001sc,Kovchegov:2001ni}
(see \cite{Jalilian-Marian:2005jf} for a review). We apply the eikonal
approximation to Einstein equations, which allows us to construct an
exact solution for the energy-momentum tensor of the produced medium
in the case of delta-function shock waves, given in \eq{Tmnd}.
(Eikonal approximation in AdS/CFT was studied before in
\cite{Brower:2007qh,Levin:2008vj,Cornalba:2007zb,Cornalba:2006xm,Cornalba:2006xk}.)
Our solution would receive energy-suppressed corrections if shock
waves of finite width are considered. We note that the energy-momentum
tensor (\ref{Tmnd}) is not that of ideal hydrodynamics, indicating
that the system does not reach isotropization/thermalization in
proton-nucleus approximation to the collision. Resumming graviton
exchanges with the nucleus shock wave to all orders allows us to
demonstrate the stopping of the proton shock wave explicitly. The
relevant component of the energy-momentum tensor of the proton in
shown in \eq{stop}: one can explicitly see that it goes to zero as the
light cone coordinate $x^+$ (in which direction the proton was
initially moving) is increasing.

We also apply the eikonal treatment to the delta-prime shock waves.
The results are quite interesting: we show that in the eikonal
approximation the series in graviton exchanges terminates at the level
of two graviton exchanges with the nucleus shock wave. Thus the NLO
result for the energy-momentum tensor is, in fact, exact for the case
of proton-nucleus collisions! The energy-momentum tensor for
delta-prime shock waves is shown in \eq{Tmnp}. It is clear from
\eq{Tmnp} that the produced medium distribution has a strong rapidity
dependence. Therefore it seems unlikely that rapidity-independent
Bjorken hydrodynamics geometry of \cite{Janik:2005zt} could result
from a collision of two shock waves in AdS$_5$ space, though indeed a
further study of the full nucleus-nucleus scattering problem is needed
to unambiguously answer this question.

We will conclude in Sect. \ref{conc} by summarizing our main results.


\section{Perturbative Expansion in Graviton Exchanges}
\label{pert}


\subsection{General Setup}

Consider a collision of two ultrarelativistic nuclei. Assume for
simplicity that the nuclei have infinite transverse extent and the
same longitudinal thickness at all impact parameters. The
energy-momentum tensors of the two nuclei can be written as $\langle
T_{1 \, --} (x^-) \rangle$ and $\langle T_{2 \, ++} (x^+) \rangle$
with the brackets $\langle \ldots \rangle$ denoting the averaging in
the nuclear wave functions and the light cone coordinates defined by
$x^\pm = (x^0 \pm x^3) / \sqrt{2}$ where $x^3$ is the collision axis.
The geometry of the collision is shown in \fig{spacetime}.

\FIGURE[h]{\includegraphics[width=7cm]{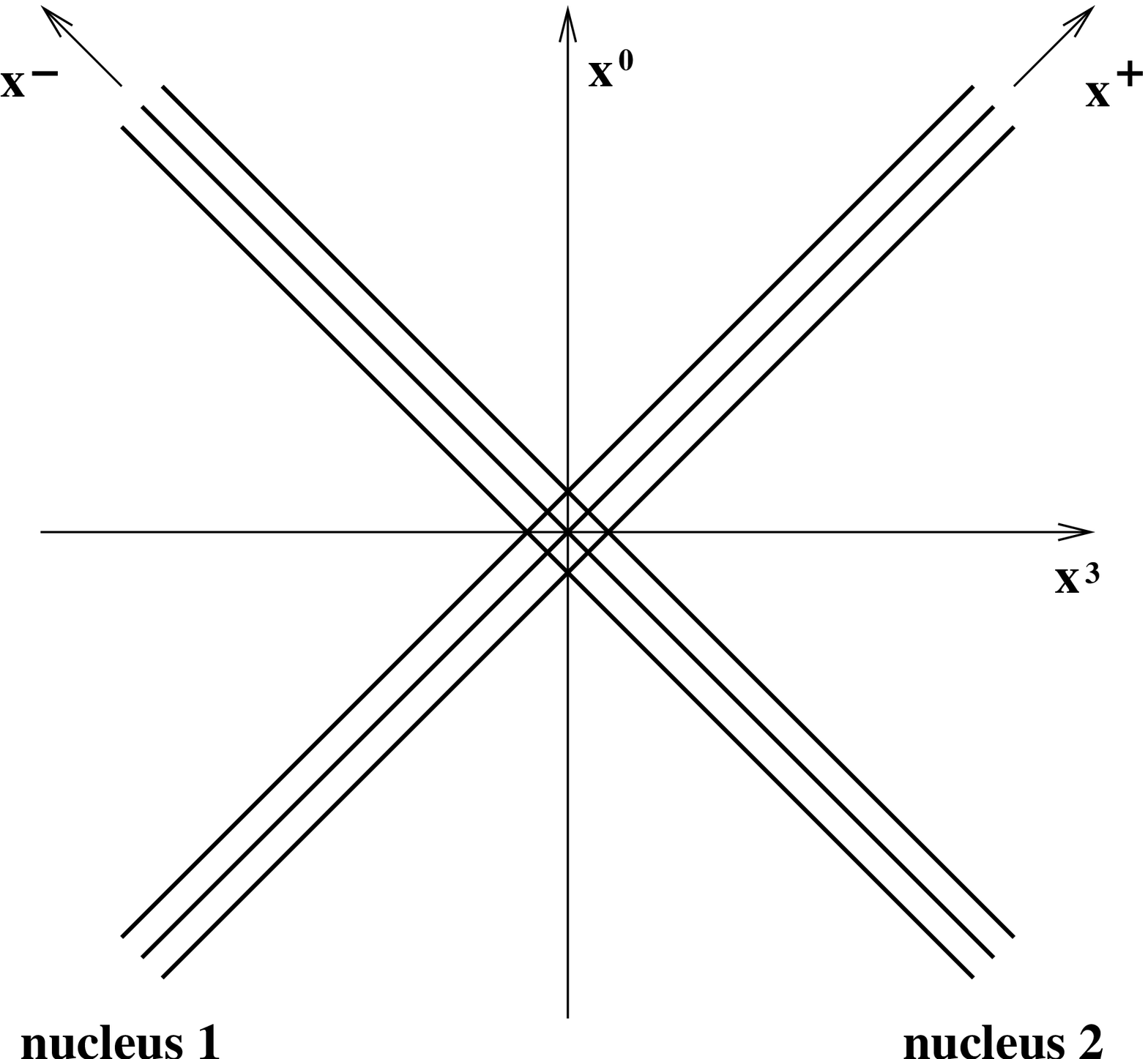}
  \caption{The space-time picture of the ultrarelativistic heavy ion 
    collision in the center-of-mass frame. The collision axis is
    labeled $x^3$, the time is $x^0$.}
  \label{spacetime}
}

As was argued in \cite{Janik:2005zt}, the geometry in AdS$_5$ dual to
each one of the nuclei in the boundary theory is given by the
following metric
\begin{align}\label{nuc1}
  ds^2 \, = \, \frac{L^2}{z^2} \, \left\{ -2 \, dx^+ \, dx^- + t_1
    (x^-) \, z^4 \, d x^{- \, 2} + d x_\perp^2 + d z^2 \right\}
\end{align}
for nucleus 1 and by 
\begin{align}\label{nuc2}
  ds^2 \, = \, \frac{L^2}{z^2} \, \left\{ -2 \, dx^+ \, dx^- + t_2
    (x^+) \, z^4 \, d x^{+ \, 2} + d x_\perp^2 + d z^2 \right\}
\end{align}
for nucleus 2. Here $d x_\perp^2 = (d x^1 )^2 + (d x^2)^2$ with $x^1$
and $x^2$ the transverse dimensions which we will denote using Latin
indices, e.g.  $x^i$. $L$ is the curvature radius of the AdS$_5$ space
and $z$ is the coordinate describing the 5th dimension with the
boundary of the AdS space at $z=0$. We have also defined
\begin{align}\label{t1}
  t_1 (x^-) \, \equiv \, \frac{2 \, \pi^2}{N_c^2} \, \langle T_{1 \,
    --} (x^-) \rangle
\end{align}
and
\begin{align}\label{t2}
  t_2 (x^+) \, \equiv \, \frac{2 \, \pi^2}{N_c^2} \, \langle T_{2 \,
    ++} (x^+) \rangle
\end{align}
in accordance with the prescription of holographic renormalization
\cite{deHaro:2000xn}. The metrics in Eqs.~(\ref{nuc1}) and
(\ref{nuc2}) are exact solutions of Einstein equations in the empty
AdS$_5$ space
\begin{align}\label{ein}
  R_{\mu\nu} + \frac{4}{L^2} \, g_{\mu\nu} = 0.
\end{align}

\begin{figure}[h]
\begin{center}
\epsfxsize=7cm
\leavevmode
\hbox{\epsffile{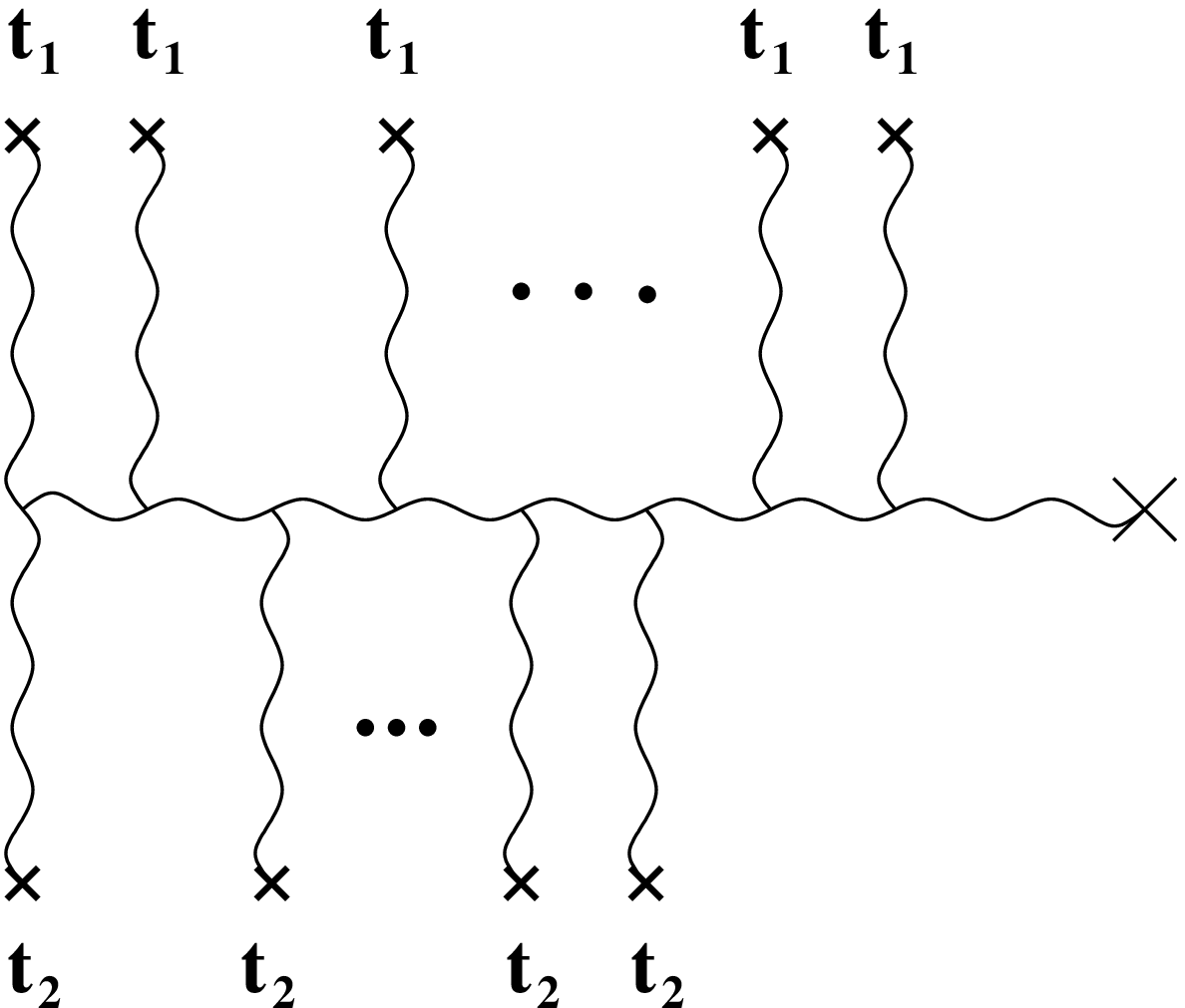}}
\end{center}
\caption{ Diagrammatic representation of the solution of classical 
  Einstein equations for the collisions of two shock waves. The wavy
  lines denote graviton exchanges between the sources at the boundary
  (thick crosses) and the bulk. The large cross denotes the point in
  the bulk where one measures the metric. The upper row of thick
  crosses denotes rescatterings in nucleus $1$, each of which
  generates a factor of $t_1$. The lower row of thick crosses denotes
  rescatterings in nucleus $2$, each of which generates a factor of
  $t_2$.}
\label{AA}
\end{figure}

Our goal is to construct the geometry in AdS$_5$ dual to the collision
of two shock waves given by Eqs.~(\ref{nuc1}) and (\ref{nuc2}). In
\cite{Albacete:2008vs} we argued that the single shock wave metric in
\eq{nuc1} (or in \eq{nuc2}) corresponds to the single-graviton
exchange between the source nucleus at the boundary and the point in
the bulk where the metric is measured. The solution of Einstein
equations (\ref{ein}) for the collision of two shock waves can
therefore be represented as a sum of tree-level graviton exchange
diagrams, as shown in \fig{AA}.  There the source nuclei are
represented by thick crosses, with nucleus $1$ given by the crosses on
the top, and nucleus $2$ given by the crosses at the bottom. As was
argued in \cite{Albacete:2008vs}, each rescattering in nucleus $1$
brings in a factor of $t_1 (x^-)$ into the metric, while each
rescattering in nucleus $2$ brings in a factor of $t_2 (x^+)$. The
large thin cross in \fig{AA} denotes the point in the bulk in the
argument of the metric, i.e., the point where the metric is
``measured''. One encounters similar diagrams but with gluons and in 4
dimensions for nuclear collisions in the framework of
McLerran-Venugopalan (MV) model
\cite{McLerran:1994vd,McLerran:1993ka,McLerran:1993ni}, as was worked
out in
\cite{Kovner:1995ts,Kovner:1995ja,Kovchegov:1998bi,Kovchegov:1997ke,Kovchegov:2000hz,Krasnitz:2003nv}.

Inspired by the graviton-exchange analogy of \fig{AA} we write the
metric dual to the full collision as \cite{Albacete:2008vs}
\begin{align}\label{2nuc2}
  ds^2 \, = \, \frac{L^2}{z^2} \, \bigg\{ -2 \, dx^+ \, dx^- + d
  x_\perp^2 + d z^2 + t_1 (x^-) \, z^4 \, d x^{- \, 2} + t_2 (x^+) \,
  z^4 \, d x^{+ \, 2} + o (t_1 \, t_2) \bigg\}.
\end{align}
Indeed the interesting unknown part of the answer is in the term
denoted $o (t_1 \, t_2)$ in \eq{2nuc2}: this term comprises all higher
order graviton exchanges, i.e., higher powers of $t_1 (x^-)$ and $t_2
(x^+)$. The first term in this expansion, the term proportional to
$t_1 \, t_2$ was found in \cite{Albacete:2008vs}. For a particular
form of $t_1 (x^-)$ and $t_2 (x^+)$ given by delta-functions, the
expansion to several higher orders in $t_1$ and $t_2$ was constructed
in \cite{Grumiller:2008va}.

To construct a series in graviton exchanges for a general form of
$t_1$ and $t_2$ and to set up the general problem we write
\begin{align}\label{2nuc_gen}
  ds^2 \, = \, \frac{L^2}{z^2} \, \bigg\{ -\left[ 2 + G (x^+, x^-, z)
  \right] \, dx^+ \, dx^- + \left[ t_1 (x^-) \, z^4 + F (x^+, x^-, z)
  \right] \, d x^{- \, 2} \notag \\ + \left[ t_2 (x^+) \, z^4 +
    {\tilde F} (x^+, x^-, z) \right] \, d x^{+ \, 2} + \left[ 1 + H
    (x^+, x^-, z) \right] \, d x_\perp^2 + d z^2 \bigg\}.
\end{align}
The unknown functions $F (x^+, x^-, z)$, ${\tilde F} (x^+, x^-, z)$,
$G (x^+, x^-, z)$, and $H (x^+, x^-, z)$ contain all higher powers of
$t_1$ and $t_2$. Note that as \eq{nuc1} and \eq{nuc2} are exact
solution of Einstein equations (\ref{ein}), the functions $F$, $\tilde
F$, $G$, and $H$ contain at least one power of $t_1$ and $t_2$ each
\cite{Albacete:2008vs}.

Substituting the metric of \eq{2nuc_gen} into Einstein equations
(\ref{ein}) yields a very complicated system of non-linear equations.
It is likely that the solution of these equations is only possible
numerically. Here we will build on the results of
\cite{Albacete:2008vs} to construct the first few steps of the
perturbative expansion: we will construct the next-to-leading order
(NLO) and the next-to-next-to-leading order (NNLO) corrections to $F$,
$\tilde F$, $G$ and $H$. NLO corrections resum terms containing $t_1^2
t_2$ and $t_1 t_2^2$, while NNLO corrections include terms with $t_1^3
t_2$, $t_1^2 t_2^2$ and $t_1 t_2^3$. (The powers of $t_1$ and $t_2$
should not be taken literally, they only indicate the number of times
$t_1$ and $t_2$ enter the expression.) In Section \ref{pA} we will use
the eikonal approximation to resum one power of $t_1$ and {\sl all}
powers of $t_2$.

Before we start the calculations let us first point out that, according
to the prescription of holographic renormalization
\cite{deHaro:2000xn}, if we expand the unknown coefficients of the
metric in \eq{2nuc_gen} into a series in powers of $z^2$
\begin{align}\label{coef_exp}
  F (x^+, x^-, z) \, = \, z^4 \, \sum\limits_{n=0}^\infty \, F_n (x^+,
  x^-) \, z^{2 \, n}, \ \ \ {\tilde F} (x^+, x^-, z) \, = \, z^4 \,
  \sum\limits_{n=0}^\infty \, {\tilde F}_n (x^+, x^-) \, z^{2 \, n}, \notag \\
  G (x^+, x^-, z) \, = \, z^4 \, \sum\limits_{n=0}^\infty \, G_n (x^+,
  x^-) \, z^{2 \, n}, \ \ \ H (x^+, x^-, z) \, = \, z^4 \,
  \sum\limits_{n=0}^\infty \, H_n (x^+, x^-) \, z^{2 \, n},
\end{align}
then the expectation value of the energy-momentum tensor of the matter
produced in the collision in the boundary theory is given by the first
coefficients in the expansion in \eq{coef_exp}:
\begin{align}\label{prod_Tmn1}
  \langle T^{++}\rangle & = \frac{N_c^2}{2 \, \pi^2} \, F_0 (x^+, x^-)
  & \langle T^{--}\rangle & = \frac{N_c^2}{2 \, \pi^2}
  \, {\tilde F}_0 (x^+, x^-) \nonumber \\
  \langle T^{+-}\rangle & = - \frac{1}{2} \, \frac{N_c^2}{2 \, \pi^2}
  \, G_0 (x^+, x^-) & \langle T^{i j}\rangle & = \frac{N_c^2}{2
    \,\pi^2} \, \delta^{i j} \, H_0 (x^+, x^-).
\end{align}
Einstein equations (\ref{ein}) impose two constraints on the
energy-momentum tensor: tracelessness
\begin{align}\label{trless}
  \langle T_\mu^{\ \mu}\rangle =0
\end{align}
and energy-momentum conservation
\begin{align}\label{emcons}
  \partial_\nu \langle T^{\mu\nu}\rangle =0.
\end{align}
Imposing the constraints (\ref{trless}) and (\ref{emcons}) on the
energy-momentum tensor in \eq{prod_Tmn1} we easily see that the
energy-momentum tensor can be expressed in terms of a single unknown
function:
\begin{align}\label{prod_Tmn2}
  \langle T^{++}\rangle & = - \frac{N_c^2}{2 \, \pi^2} \,
  \frac{\partial_{-}}{\partial_{+}} \, H_0 (x^+, x^-) & \langle
  T^{--}\rangle & = - \frac{N_c^2}{2 \, \pi^2}
  \, \frac{\partial_{+}}{\partial_{-}} \, H_0 (x^+, x^-) \nonumber \\
  \langle T^{+-}\rangle & = \frac{N_c^2}{2 \, \pi^2} \, H_0 (x^+, x^-)
  & \langle T^{i j}\rangle & = \frac{N_c^2}{2 \,\pi^2} \, \delta^{i j}
  \, H_0 (x^+, x^-).
\end{align}
Here we defined the following integrations
\begin{align}\label{ints}
  \frac{1}{\partial_{+}} [\ldots](x^+) \, \equiv \,
  \int\limits_{-\infty}^{x^+} \, d x'^+ \, [\ldots](x'^+), \ \ \ 
  \frac{1}{\partial_{-}} [\ldots](x^-) \, \equiv \,
  \int\limits_{-\infty}^{x^-} \, d x'^- \, [\ldots](x'^-).
\end{align}

Eqs. (\ref{prod_Tmn2}) demonstrate that only one metric coefficient in
(\ref{2nuc_gen}) is needed to construct the energy-momentum tensor of
the produced matter in the boundary theory.


\subsection{NLO Results}
\label{NLO}


\subsubsection{NLO Calculation}

To systematically include the graviton exchanges of \fig{AA} into the
metric of \eq{2nuc_gen} we expand the coefficients of the metric in
powers of $t_1$ and $t_2$. We start by writing
\begin{align}\label{texp}
  F (x^+, x^-, z) \, & = \, F^{(0)} (x^+, x^-, z) + F^{(1)} (x^+, x^-,
  z) + F^{(2)} (x^+, x^-, z) + \ldots \notag \\
  {\tilde F} (x^+, x^-, z) \, & = \, {\tilde F}^{(0)} (x^+, x^-, z) +
  {\tilde F}^{(1)} (x^+, x^-,
  z) + {\tilde F}^{(2)} (x^+, x^-, z) + \ldots \notag \\
  G (x^+, x^-, z) \, & = \, G^{(0)} (x^+, x^-, z) + G^{(1)} (x^+, x^-,
  z) + G^{(2)} (x^+, x^-, z) + \ldots \notag \\
  H (x^+, x^-, z) \, & = \, H^{(0)} (x^+, x^-, z) + H^{(1)} (x^+, x^-,
  z) + H^{(2)} (x^+, x^-, z) + \ldots
\end{align}
where the superscript ${(0)}$ denotes terms containing $t_1 \, t_2$,
i.e., quadratic in $t$'s, the superscript $(1)$ denotes terms cubic in
$t$'s (i.e., terms containing $t_1^2 \, t_2$ and $t_1 \, t_2^2$), the
superscript $(2)$ denotes terms quadric in $t$'s, etc. Note that the
expansion in $t$'s in \eq{texp} is independent of the expansion in
$z$'s in \eq{coef_exp}: each term in the expansion in \eq{texp} can in
turn be expanded in powers of $z^2$ as was done in \eq{coef_exp}, and
vice versa.

\FIGURE{\includegraphics[width=3cm]{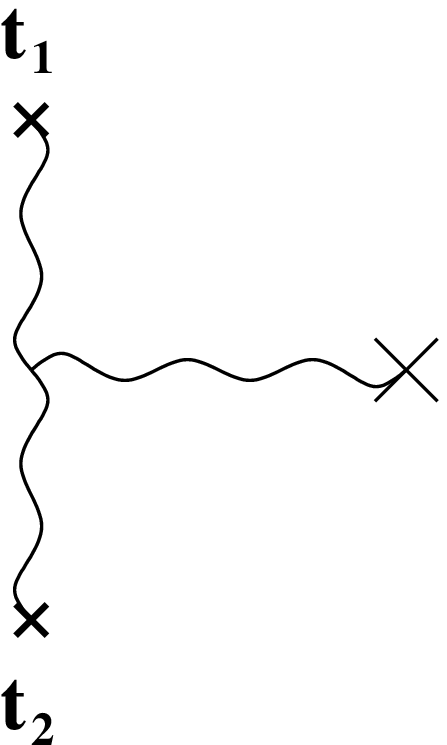}
  \caption{Graviton diagram corresponding to the LO solution of 
    Einstein equations found in \cite{Albacete:2008vs}.}
  \label{LOfig}
}

The leading order (LO) terms in \eq{texp} denoted by the superscript
${(0)}$ were found in \cite{Albacete:2008vs}. For completeness let us
quote the results here:
\begin{align}\label{LO}
  F^{(0)} (x^+, x^-, z) \, & = \, - \lambda_1 (x^+, x^-) \, z^4 -
  \frac{1}{6} \, \partial_-^2 h_0 (x^+, x^-) \, z^6 - \frac{1}{16} \,
  \partial_-^2
  h_1 (x^+, x^-) \, z^8 \notag \\
  {\tilde F}^{(0)} (x^+, x^-, z) \, & = \, - \lambda_2 (x^+, x^-) \,
  z^4 - \frac{1}{6} \, \partial_+^2 h_0 (x^+, x^-) \, z^6 -
  \frac{1}{16} \, \partial_+^2
  h_1 (x^+, x^-) \, z^8 \notag \\
  G^{(0)} (x^+, x^-, z) \, & = \, - 2 \, h_0 (x^+, x^-) \, z^4 - 2 \,
  h_1
  (x^+, x^-) \, z^6 + \frac{2}{3} \, t_1 (x^-) \, t_2 (x^+) \, z^8 \notag \\
  H^{(0)} (x^+, x^-, z) \, & = \, h_0 (x^+, x^-) \, z^4 + h_1 (x^+,
  x^-) \, z^6.
\end{align}
We defined \cite{Albacete:2008vs}
\begin{align}\label{LOstuff}
  h_0 (x^+, x^-) \, & = \, \frac{8}{\partial_+^2 \, \partial_-^2} \,
  t_1 (x^-) \, t_2 (x^+), \ \ \ h_1 (x^+, x^-) \, = \, \frac{4}{3 \,
    \partial_+ \, \partial_-} \, t_1 (x^-) \, t_2 (x^+) \notag \\
  \lambda_1 (x^+, x^-) \, & = \, \frac{\partial_{-}}{\partial_{+}} \,
  h_0 (x^+, x^-), \ \ \ \lambda_2 (x^+, x^-) \, = \,
  \frac{\partial_{+}}{\partial_{-}} \, h_0 (x^+, x^-).
\end{align}
The diagram corresponding to the LO solution given by Eqs. (\ref{LO})
and (\ref{LOstuff}) is shown in \fig{LOfig}. Note that the $z^4$ terms
in Eq. (\ref{LO}) adhere to the pattern outlined in Eqs.
(\ref{prod_Tmn2}).

To find the NLO terms denoted by superscript $(1)$ in \eq{texp} we
substitute the metric (\ref{2nuc_gen}) with the coefficients expanded
according to \eq{texp} into Einstein equations (\ref{ein}). Expanding
the Einstein equations to the cubic order in $t$'s yields the
following equations for $G^{(1)}$ and $H^{(1)}$
\begin{subequations}\label{ein1}
\begin{align}
  (\bot \bot) \hspace*{0.15in}& G^{(1)}_{z} + 5 \, H^{(1)}_{z} - z \,
  H^{(1)}_{z \, z} + 2 \, z \, H^{(1)}_{x^{+} \, x^{-}} + \delta_{7}
  \,
  z^{7} + \delta_{9} \, z^{9} + \delta_{11} \, z^{11} = 0 \label{cmpactpp} \\
  (zz) \hspace*{0.15in}& G^{(1)}_{z} + 2 H^{(1)}_{z} - z \, G^{(1)}_{z
    \, z} - 2 \, z \, H^{(1)}_{z \, z} + 48 \, \alpha_{7} \, z^{7} +
  80 \, \alpha_{9} \, z^{9} + 120 \, \alpha_{11}\, z^{11} = 0.
  \label{cmpactzz}
\end{align}
\end{subequations}
The coefficients $\delta_{7}$, $\delta_{9}$, $\delta_{11}$,
$\alpha_{7}$, $\alpha_{9}$, $\alpha_{11}$ are known functions of $t_1$
and $t_2$ the exact form of which is not important here. The
subscripts $z$, $x^+$ and $x^-$ indicate partial derivatives with
respect to these variables. Eqs. (\ref{cmpactpp}) and (\ref{cmpactzz})
are labeled according to the lowercase Einstein equations components.

Solving \eq{cmpactpp} for $G^{(1)}_{z}$ and substituting the result
into \eq{cmpactzz} yields the following equation for $H^{(1)}$
\begin{align}\label{cmpactppzz}
  - 3 \, H^{(1)}_{z} + 3 \, z \, H^{(1)}_{z \, z} - z^{2} \,
  H^{(1)}_{z \, z \, z} + 2 \, z^{2} \, H^{(1)}_{x^{+} \, x^{-} \, z}
  + 12 \, \psi_{7} \, z^{7} + 16 \, \psi_{9} \, z^{9} + 20 \,
  \psi_{11} \, z^{11} = 0
\end{align}
with the coefficients $\psi$ given by
\begin{align}\label{ps}
  & \psi_7 \, = \, \frac{4}{3} \, \left[ t_{2}(x^+) \,
    \lambda_1(x^+,x^-) + t_{1}(x^-) \, \lambda_2(x^+,x^-) \right] \notag \\
  & \psi_9 \, =  \, \frac{3}{4} \, \left[ t_{2}(x^+) \, h_{0 \ x^- \, x^-}(x^+, x^-) 
    + t_{1}(x^-) \, h_{0 \ x^+ \, x^+} (x^+,x^-) \right] \notag \\
  & \psi_{11} \, = \, \frac{3}{5} \, \left[ t_{2}(x^+) \, h_{1 \ x^-
      \, x^-}(x^+,x^-) + t_{1}(x^-) \, h_{1 \ x^+ \, x^+}(x^+,x^-)
  \right].
\end{align}

To find the solution of \eq{cmpactppzz} we follow the strategy used in
\cite{Albacete:2008vs}. We first expand $H^{(1)}$ into a series in
powers of $z^2$
\begin{align}\label{H1}
  H^{(1)} (x^+, x^-, z) \, = \, z^4 \, \sum\limits_{n=0}^\infty \,
  H_n^{(1)} (x^+, x^-) \, z^{2 \, n}.
\end{align}
Substituting \eq{H1} into \eq{cmpactppzz} and requiring that the
coefficients at each power of $z$ on the left hand side are zero
yields the recursion relation
\begin{align}\label{H3}
  H^{(1)}_n \, (-2n) \, (2+n) + H^{(1)}_{n-1; \ x^+x^-} + \psi_7 \,
  \delta_{n,2} + \psi_9 \, \delta_{n,3} + \psi_{11} \, \delta_{n,4} =
  0 \hspace{0.4in} n\geq1.
\end{align}
Arguing just like in \cite{Albacete:2008vs} that causality requires
the series (\ref{H1}) to terminate at some finite order, we see that
the series can only be terminated if $H^{(1)}_4 =0$. The solution of
\eq{cmpactppzz} is thus given by
\begin{align}\label{H}
  H^{(1)} (x^+,x^-,z) \, = \, H^{(1)}_0 (x^+,x^-) \, z^4 +
  H^{(1)}_1(x^+,x^-) \, z^6 + H^{(1)}_2 (x^+,x^-) \, z^8 + H^{(1)}_3
  (x^+,x^-) \, z^{10}
\end{align}
with the coefficients
\begin{subequations}\label{H1coefs}
\begin{align}\label{H10}
  H^{(1)}_0 & = - \frac{6}{(\partial_{+} \, \partial_{-})^{2}} \,
  \psi_{7} - \frac{96}{(\partial_{+} \, \partial_{-})^{3}} \, \psi_{9}
  - \frac{2880}{(\partial_{+} \, \partial_{-})^{4}} \, \psi_{11} \\
  H^{(1)}_{1} & = - \frac{1}{\partial_{+} \, \partial_{-}} \, \psi_{7}
  - \frac{16}{(\partial_{+} \, \partial_{-})^{2}} \, \psi_{9} -
  \frac{480}{(\partial_{+} \, \partial_{-})^{3}} \, \psi_{11} \\
  H^{(1)}_{2} & = - \frac{1}{\partial_{+} \, \partial_{-}} \, \psi_{9}
  -
  \frac{30}{(\partial_{+} \, \partial_{-})^{2}} \, \psi_{11} \\
  H^{(1)}_{3} & = - \frac{1}{\partial_{+} \, \partial_{-}} \,
  \psi_{11}.
\end{align}
\end{subequations}
Using $H^{(1)}$ from \eq{H} in \eq{cmpactpp} one can easily find
$G^{(1)}$. With the help of two other components of Einstein equations
which are not shown here explicitly we can find (and have found) $F
^{(1)}$ and ${\tilde F}^{(1)}$. The remaining components of Einstein
equations do not generate further constraints.

\FIGURE{\includegraphics[width=7cm]{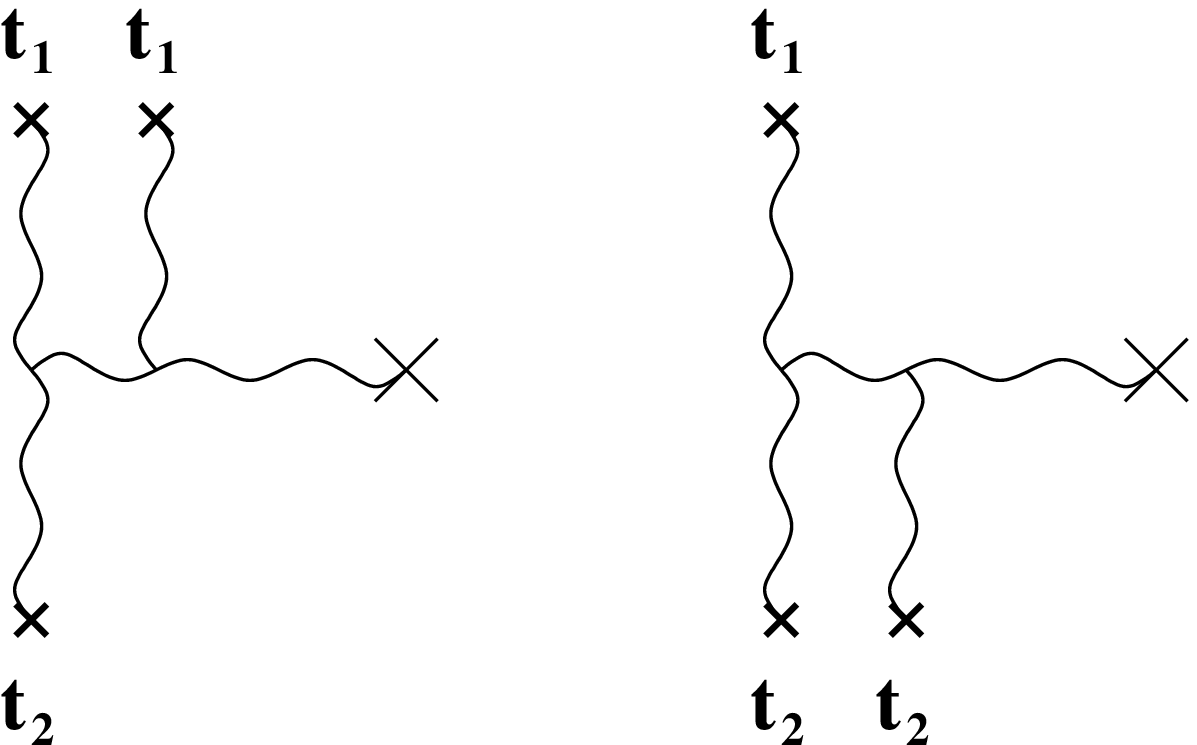}
  \caption{Graviton diagrams corresponding to the NLO solution of 
    Einstein equations found in this Section.}
  \label{NLOfig}
}

Note that $G^{(1)}$, $F ^{(1)}$ and ${\tilde F}^{(1)}$ are indeed
needed to construct the metric at higher orders in the expansion in
$t$'s. However, as we argued above and as shown in \eq{prod_Tmn2},
only $H^{(1)}_0$ is needed to obtain the energy-momentum tensor of the
produced matter at NLO. Since at NLO $G^{(1)}$, $F ^{(1)}$ and
${\tilde F}^{(1)}$ are not needed for the boundary theory physics that
we are interested in here, we will not present explicit expressions
for these quantities.

The NLO solution in Eqs. (\ref{H}) and (\ref{H1coefs}) is represented
diagrammatically in terms of graviton exchanges in \fig{NLOfig}. 
As shown in \fig{NLOfig} the NLO solution consists of a single
rescattering in one nucleus and a double rescattering in another
nucleus. As can be seen from Eqs. (\ref{H1coefs}) and (\ref{ps}), NLO
solution includes terms with two powers of $t_1$ and one power of
$t_2$ and terms with two powers of $t_2$ and one power of $t_1$.


\subsubsection{Delta-Function Shock Waves at NLO}
\label{NLOdeltas}

It is instructive to find what the obtained results give for specific
shock waves described by particular forms of $t_1 (x^-)$ and $t_2
(x^+)$. Define the transverse pressure $p$ of the produced medium by
\begin{align}\label{pdef}
  \langle T^{i\, j} \rangle = \delta^{ij} \, p. 
\end{align}
Combining \eq{pdef} with Eqs. (\ref{prod_Tmn2}) and (\ref{texp})
yields
\begin{align}\label{pexp}
  p (x^+, x^-) \, = \, \frac{N_c^2}{2 \, \pi^2} \, \left[ H^{(0)}_0
    (x^+, x^-) + H^{(1)}_0 (x^+, x^-) + H^{(2)}_0 (x^+, x^-) + \ldots
  \right].
\end{align}
Let us for simplicity concentrate on this component of the
energy-momentum tensor: all others can be also easily constructed
using \eq{prod_Tmn2}.

Following the original suggestion of \cite{Janik:2005zt} (see also
\cite{Grumiller:2008va}) let us first consider delta-function shock
waves
\begin{align}\label{deltas}
  t_1 (x^-) = \mu_1 \, \delta (x^-), \ \ \ t_2 (x^+) = \mu_2 \, \delta
  (x^+).
\end{align}
As was argued in \cite{Albacete:2008vs}, the delta-function shock
waves give a solution of Einstein equations having correct qualitative
features of the solution for any colliding shock waves with
non-negative $t_1$ and $t_2$. However, plugging the delta-functions
from (\ref{deltas}) into \eq{LOstuff} and then into \eq{ps} we
immediately encounter a problem: we obtain products of delta-functions
and theta functions, like $\delta (x^+) \, \theta (x^+)$. To properly
handle those terms let us regulate the delta-functions by spearing
them along the light cone directions:
\begin{align}\label{thetas}
  t_1 (x^-) = \frac{\mu_1}{a_1} \, \theta (x^-) \, \theta (a_1 - x^-),
  \ \ \ t_2 (x^+) = \frac{\mu_2}{a_2} \, \theta (x^+) \, \theta (a_2 -
  x^+).
\end{align}
To be more specific let us consider in the boundary theory a collision
of two ultrarelativistic nuclei with large light-cone momenta per
nucleon $p_1^+$, $p_2^-$, and atomic numbers $A_1$ and $A_2$. In order
to avoid $N_c^2$ suppression in each graviton exchange coming from the
Newton's constant (see e.g. \eq{t1}) let us assume that each nucleon
in the nucleus has $N_c^2$ nucleons in it. This factor of $N_c^2$ in
$\langle T_{1 --} \rangle$ and $\langle T_{2 ++} \rangle$ cancels the
factor of $1/N_c^2$ in Eqs. (\ref{t1}) and (\ref{t2}). In the end one
obtains, similar to \cite{Albacete:2008vs,Albacete:2008ze}
\begin{align}\label{mus}
  \mu_{1} \sim p_{1}^+ \, \Lambda_1^2 \, A_1^{1/3}, \ \ \ \mu_{2} \sim
  p_{2}^- \, \Lambda_2^2 \, A_2^{1/3},
\end{align}
while the Lorentz-contracted widths of the nuclei are
\begin{align}\label{as}
  a_1 \sim \frac{A_1^{1/3}}{p_1^+}, \ \ \ a_2 \sim
  \frac{A_2^{1/3}}{p_2^-}.
\end{align}
The scales $\Lambda_1$ and $\Lambda_2$ are the typical transverse
momentum scales describing the two nuclei \cite{Albacete:2008vs},
similar to the saturation scales.

Using \eq{thetas} along with Eqs. (\ref{pexp}), (\ref{LO}),
(\ref{LOstuff}), (\ref{H}), and (\ref{H10}), we obtain
\begin{align}\label{pNLO}
  p (x^+, x^-) = \frac{N_c^2}{2 \, \pi^2} \, 8 \, \mu_1 \, \mu_2 \,
  x^+ \, x^- \, \theta (x^+) \, \theta (x^-) \, \left[ 1 - 12 \, \mu_1
    \, x^+ \, (x^-)^2 - 12 \, \mu_2 \, (x^+)^2 \, x^- + \ldots
  \right].
\end{align}
In arriving at \eq{pNLO} we neglected terms suppressed by powers of
$a_1 / x^-$ and $a_2 / x^+$. Thus \eq{pNLO} is only valid when 
\begin{align}\label{bounds1}
  \frac{a_1}{x^-} \ll 1, \ \ \  \frac{a_2}{x^+} \ll 1.
\end{align}
However this is not the only constraint on applicability of \eq{pNLO}:
requiring that $o (a_1 / x^-, \, a_2 / x^+)$ corrections to the NLO
terms are much smaller than LO terms, employing Eqs. (\ref{mus}) and
(\ref{as}), and assuming for simplicity that $p_1^+ \sim p_2^-$,
$\Lambda_1 \approx \Lambda_2 \equiv \Lambda$, $A_1 \approx A_2 \equiv
A$, we obtain another restriction
\begin{align}\label{bounds2}
  \Lambda \, A^{1/3} \, \tau \ll 1
\end{align}
with the proper time $\tau = \sqrt{2 \, x^+ \, x^-}$.  Hence \eq{pNLO}
is valid at relatively early proper times and acquires order-1
corrections at later times. Indeed \eq{pNLO} provides an exact
solution in the formal limit of $a_1, a_2 \rightarrow 0$ which reduces
$t_1$ and $t_2$ back to the delta-function expressions given in
\eq{deltas}.  However, Eqs. (\ref{mus}) and (\ref{as}) demonstrate
that if we keep track of the physical origin of the delta-functions,
the infinitely-thin nucleus limit gets more involved. Of course one
can always postulate the nuclei to be very thin in the longitudinal
direction while keeping their atomic numbers fixed, thus making the
formal $a_1, a_2 \rightarrow 0$ limit possible: such limit is not
attainable in real life, but it is a mathematically well-defined
procedure.

\eq{pNLO} agrees with the appropriate result obtained in
\cite{Grumiller:2008va} for delta-function shock waves.


\subsubsection{Delta-Prime Shock Waves at NLO}

In \cite{Albacete:2008vs} it was argued that delta-function shock
waves considered in Sect. \ref{NLOdeltas} come to a complete stop
shortly after the collision, possibly leading to a formation of a
black hole. For the boundary theory this implied that the colliding
nuclei stop after the collision and thermalize leading to Landau-like
hydrodynamics \cite{Landau:1953gs}. This scenario would lead to strong
baryon stopping in the collisions, which is not what is observed by
the experiments at RHIC. Combined with the many successes of
small-coupling based approaches in describing RHIC data sensitive to
early-time dynamics (for a review see \cite{Jalilian-Marian:2005jf}),
this led us to conclude that one can not adequately describe entire
heavy ion collision within a strong coupling framework. Thus
collisions of delta-function shock waves in AdS$_5$ are not relevant
for the heavy ion collisions, in which it is very likely that the
initial stages of the collisions are weakly-coupled. In
\cite{Albacete:2008vs} to try to mimic these weak coupling effects we
suggested using unphysical delta-prime shock waves
\begin{align}\label{tzero}
  t_1 (x^-) \, = \, \Lambda_1^2 \, A_1^{1/3} \, \delta' (x^-), \ \ \ 
  t_2 (x^+) \, = \, \Lambda_2^2 \, A_2^{1/3} \, \delta' (x^+).
\end{align}
The shock waves in \eq{tzero} are fundamentally different from those
in Sect. \ref{NLOdeltas} as the integrals of these shock wave profiles
over all $x^-$'s and/or $x^+$'s give zero. The shock waves
(\ref{tzero}) have unphysical energy-density on the light cone.
However, in the LO calculations carried out in \cite{Albacete:2008vs}
it was shown that the behavior of the produced matter in the forward
light cone of a collision of two shock waves (\ref{tzero}) gives a
well-behaved physical distribution of matter. This should be
contrasted with the physical shock waves in Sect. \ref{NLOdeltas}, for
which, due to nuclear stopping, the remnants of the colliding nuclei
would deviate from their initial light cone trajectories and drift
into the forward light cone.

To use the shock waves of \eq{tzero} for calculating the NLO
contribution to the transverse pressure $p$ we have to regulate them.
We do that by rewriting (\ref{tzero}) as
\cite{Kovchegov:1996ty,Kovchegov:1997ke}
\begin{align}\label{tzero_reg}
  t_1 (x^-) \, = \, \Lambda_1^2 \, \sum\limits_{i=1}^{A_1^{1/3}} \,
  \delta' (x^- - x^-_i)
  \notag \\
  t_2 (x^+) \, = \, \Lambda_2^2 \, \sum\limits_{i=1}^{A_2^{1/3}} \,
  \delta' (x^+ - x_i^+).
\end{align}
Each delta-prime in \eq{tzero_reg} corresponds to a thin slice of a
shock wave (a ``nucleon'') localized around the longitudinal
coordinate $x^\pm_i$. The coordinates $x^-_i$ are localized to the
interval $[0,a_1]$ of the $x^-$ axis, while the coordinate $x^+_i$ are
localized to the interval $[0,a_2]$ of the $x^+$ axis.

Employing \eq{tzero_reg} in Eqs. (\ref{pexp}), (\ref{LO}),
(\ref{LOstuff}), (\ref{H}), and (\ref{H10}), and assuming that $A_1,
A_2 \gg 1$, yields the transverse pressure 
\begin{align}\label{ppNLO}
  p (x^+, x^-) \, = \, \frac{N_c^2}{2 \, \pi^2} \, 8 \, \Lambda_1^2 \,
  A_1^{1/3} \, \Lambda_2^2 \, A_2^{1/3} \, \theta (x^+) \, \theta
  (x^-) \, \left\{ 1 - 40 \, \left[ \Lambda_1^2 \, A_1^{1/3} +
      \Lambda_2^2 \, A_2^{1/3} \right] \, x^+ \, x^- \right. \notag \\
  \left. - 36 \, \left[ \Lambda_1^2 \, A_1^{1/3} \, p_1^+ \, x^+ \,
      (x^-)^2 + \Lambda_2^2 \, A_2^{1/3} \, p_2^- \, (x^+)^2 \, x^-
    \right] + \ldots \right\}.
\end{align}
\eq{ppNLO} is derived in Appendix \ref{A}.  Just like with \eq{pNLO},
in arriving at \eq{ppNLO} we have neglected terms suppressed by
additional powers of energy, i.e., we assumed the condition
(\ref{bounds1}) to be valid.  At the same time we did not have to
assume that the bound (\ref{bounds2}) applies.

From \eq{ppNLO} we see that NLO corrections in the transverse pressure
are of two types: they can be rapidity/energy-independent, like the
second term in the square brackets, which is proportional to $x^+ \,
x^- \sim \tau^2$. They can also be rapidity/energy-dependent, like the
last term in the square brackets in \eq{ppNLO}, which is proportional
to, say, $(x^+)^2 \, x^- \sim \tau^3 \, e^{\eta}$, where we defined
the space-time rapidity $\eta = (1/2) \ln (x^+ / x^-)$. That term also
includes explicit powers of the large momentum components $p_1^+$ and
$p_2^-$, i.e., it is explicitly energy-dependent. Indeed if $p_1^+ \,
x^- \gg 1$ or $p_2^- \, x^+ \gg 1$ the last term in the square
brackets of \eq{ppNLO} dominates over the second term in the brackets.


\subsection{NNLO Results}
\label{NNLOsec}

Evaluation of the NNLO terms goes along the same lines as the NLO
calculation. One plugs the expansion of \eq{texp} into Einstein
equations (\ref{ein}) and expands the resulting equations up to the
quadric order in $t$'s. In particular one obtains the following
equations for $G^{(2)}$ and $H^{(2)}$
\begin{subequations}\label{ein2}
\begin{align}
  (\bot \bot) \hspace{0.15in} G^{(2)}_{z} + 5 \, H^{(2)}_{z} - z \, 
H^{(2)}_{z \, z} + 2 \, z \, H^{(2)}_{x^{+} \, x^{-}} + \Delta_{7} \, z^{7} 
+ \Delta_{9} \, z^{9}+ \Delta_{11} \, z^{11} + \Delta_{13} \, z^{14} 
+ \Delta_{15} \, z^{15} = 0 \label{Cmpactpp} \\
  (zz) \hspace{0.15in}  G^{(2)}_{z} + 2 \, H^{(2)}_{z} - z \, G^{(2)}_{zz} 
- 2 \, z \, H^{(2)}_{z \, z} 
  + 48 \, A_{7} \, z^{7} + 80 \, A_{9} \, z^{9} + 120 \, A_{11} \, z^{11} 
+ 168 \, A_{13} \, z^{13} \notag \\ + 224 \, A_{15} \, z^{15} = 0\label{Cmpactzz}
\end{align}
\end{subequations}
with $\Delta$'s and $A$'s being some known functions of $t_1$ and
$t_2$. Eliminating $G^{(2)}$ from Eqs. (\ref{Cmpactpp}) and
(\ref{Cmpactzz}) yields
\begin{align}
  - 3 \, H^{(2)}_{z} + 3 \, z \, H^{(2)}_{z \, z} - z^{2} \,
  H^{(2)}_{z \, z \, z} + 2 \, z^{2} \, H^{(2)}_{x^{+} \, x^{-} \, z}
  + 12 \, \Psi_{7} \, z^{7} + 16 \, \Psi_{9} \, z^{9} + 20 \,
  \Psi_{11} \, z^{11} + 24 \, \Psi_{13} \, z^{13} \notag \\ + 28 \,
  \Psi_{15} \, z^{15} = 0 \label{Cmpactppzz}
\end{align}
\FIGURE[th]{\includegraphics[width=11cm]{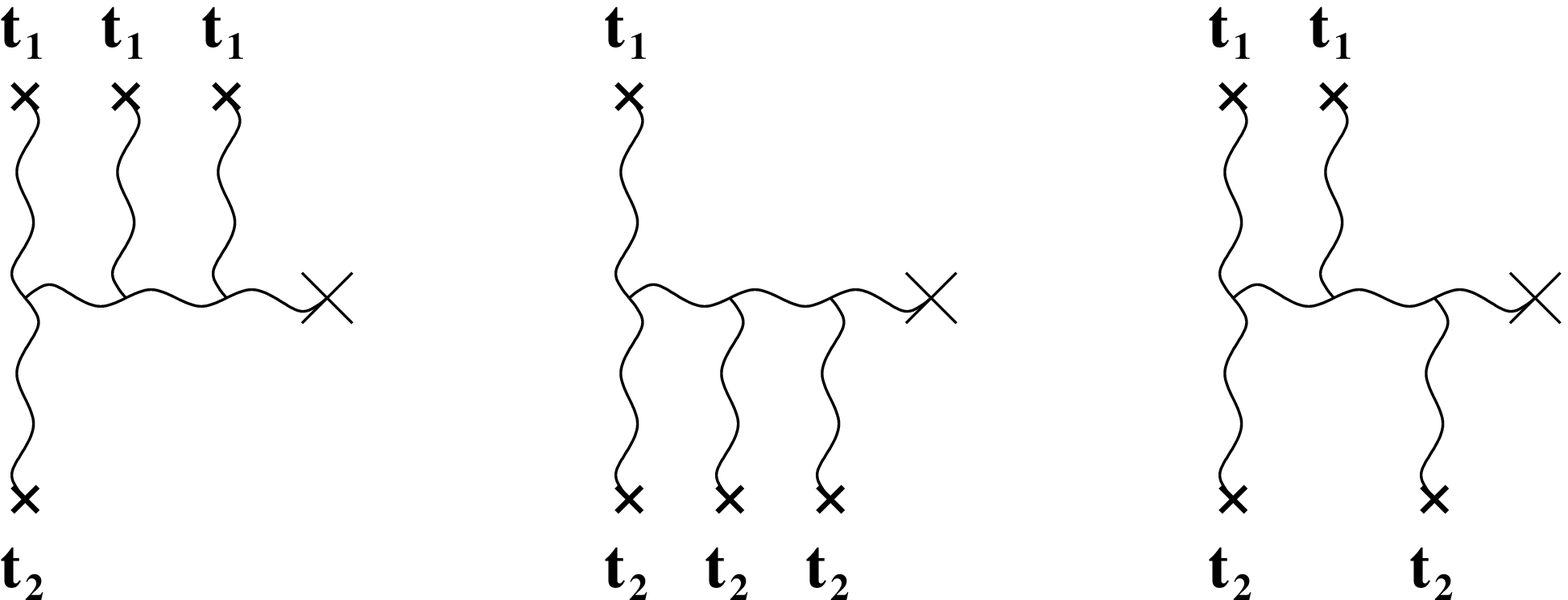}
  \caption{Some of the graviton diagrams corresponding to the NNLO solution of 
    Einstein equations found in Section \ref{NNLOsec}.}
  \label{NNLOfig}
}
with
\begin{subequations}\label{psis}
\begin{align}
  \Psi_{7}\, =\,&\frac{4}{3}\, \Big[4 \, h_0^2 - \lambda_1 \, \lambda_2 + t_1 \, \frac{\partial_+}{\partial_-} \, H_0^{(1)} + t_2 \, \frac{\partial_-}{\partial_+} \, H_0^{(1)} \Big]\\
  \Psi_{9}\, =\, &\frac{1}{4}\,\Big[-4 \, h_{0x^-} \, h_{0x^+} + 16 \,
  h_0 \, h_{0 \, x^+ \, x^-}+ \left\{3 \,
    (-\lambda_2 \,h_{0 \, x^- \, x^-} + t_2 \, H^{(1)}_{0 \, x^- \, x^-}) + (x^{+} \leftrightarrow x^{-}; \, 1\leftrightarrow 2) \right\} \Big]\\
  \Psi_{11}\, =\,&\frac{1}{60}\,\Big[ 128 \, h_0 \, t_1 \, t_2 + 34 \,
  (h_{0 \, x^+ \, x^-})^{2}-13 \, h_{0 \, x^+ \, x^+} \, h_{0 \, x^- \,
    x^-}\notag\\&
  +\left\{36 \, t_2 \, H^{(1)}_{1\, x^-\, x^-} - 10 \, h_{0\, x^+} \, h_{0 \, x^+ \, x^- \, x^-}-6 \, \lambda_{2} \, h_{0 \, x^+ \, x^- \, x^- \, x^-} + (x^+\leftrightarrow x^{-}; \, 1\leftrightarrow2) \right\}\Big]\\
  \Psi_{13}\, =\,&\frac{1}{576}\,\Big[768 \, t_1 \, t_2 \, h_{0 \, x^+
    \, x^-} -16 \, h_{0 \, x^+\, x^-\, x^-} \, h_{0 \, x^+\, x^+\,
    x^-} + \Big\{ 136 \, t_1 \, t'_{2} \, h_{0\, x^-} \notag\\& \left.
    +320 \, t_2 \, H^{(1)}_{2 \, x^- \, x^-} - 13 \, h_{0 \, x^+ \, x^- \, x^- \, x^-} \, h_{0 \, x^+ \, x^+}  + (x^{+}\leftrightarrow x^{-}; \, 1\leftrightarrow2) \right\}\Big]\\
  \Psi_{15}\, = \, & \frac{1}{504} \, \Big[ 368 \, t_1^2 \, t_2^2 -
  h_{0\, x^+\, x^-\, x^-\, x^-} \, h_{0\, x^+\, x^+\, x^+\, x^-} +
  \left\{ 270 \, t_2 \, H^{(1)}_{3\, x^-\, x^-} + 19 \, t_1 \, t'_2 \,
    h_{0 \, x^+\, x^-\, x^-} \right. \notag\\& + (x^{+}\leftrightarrow
  x^{-};\, 1\leftrightarrow2) \Big\} \Big].
\end{align}
\end{subequations}
The prime in $t'_1 (x^-)$ and in $t_2' (x^+)$ indicates derivatives
with respect to the only argument of the functions.

To find a causal solution of \eq{Cmpactppzz} one expands $H^{(2)}$
into a series in $z^2$, matches the coefficients of the powers of
$z^2$ and requires the series to terminate at some finite order to
find the coefficients. The answer then reads
\begin{align}\label{H2}
  H^{(2)} (x^+,x^-,z) \, = \, H^{(2)}_0 (x^+,x^-) \, z^4 +
  H^{(2)}_1 (x^+,x^-) \, z^6 + H^{(2)}_2 (x^+,x^-) \, z^8 + 
  H^{(2)}_3 (x^+,x^-) \, z^{10} \notag\\
  + H^{(2)}_4 (x^+,x^-) \, z^{12} + H^{(2)}_5 (x^+,x^-) \, z^{14}
\end{align}
with
\begin{subequations}\label{H22}
\begin{align}
  H^{(2)}_{0} \, & = \, \frac{6}{\partial_+ \, \partial_-} \, H^{(2)}_1 \label{Wo} \\
  H^{(2)}_{1} \, & = \, - \frac{1}{\partial_+ \, \partial_-} \,
  \psi_{7} -  \frac{16}{(\partial_+ \, \partial_-)^2} \, \Psi_{9} - \frac{(16)(30)}{(\partial_{+} \, \partial_{-})^{3}} \, \Psi_{11} - \frac{(16)(30)(48)}{(\partial_{+} \, \partial_{-})^{4}} \, \Psi_{13} - \frac{(16)(30)(48)(70)}{(\partial_{+} \, \partial_{-})^{5}} \, \Psi_{15} \\
  H^{(2)}_{2} \, & = \, -\frac{1}{(\partial_{+} \, \partial_{-})} \, \Psi_{9} - \frac{30}{(\partial_{+} \, \partial_{-})^{2}} \, \Psi_{11} - \frac{(30)(48)}{(\partial_{+} \, \partial_{-})^{3}} \, \Psi_{13} - \frac{(30)(48)(70)}{(\partial_{+}\, \partial_{-})^{4}} \, \Psi_{15} \\
  H^{(2)}_{3} \, & = \, - \frac{1}{(\partial_{+} \, \partial_{-})} \, \Psi_{11} -  \frac{48}{(\partial_{+} \, \partial_{-})^{2}} \, \Psi_{13} - \frac{(48)(70)}{(\partial_{+} \, \partial_{-})^{3}} \, \Psi_{15} \\
  H^{(2)}_{4} \, & = \, - \frac{1}{(\partial_{+} \, \partial_{-})} \,
  \Psi_{13} - \frac{70}{(\partial_{+} \, \partial_{-})^{2}} \,
  \Psi_{15} \\
  H^{(2)}_{5} \, & = \, -\frac{1}{(\partial_{+} \, \partial_{-})} \,
  \Psi_{15}.
\end{align}
\end{subequations}

Using Eqs. (\ref{H2}) with (\ref{H22}) in the remaining Einstein
equations allows one to find the other components of the metric at the
same order: $G^{(2)}$, $F^{(2)}$, and ${\tilde F}^{(2)}$. The
essential classes of diagrams resummed at NNLO are shown in
\fig{NNLOfig}. They involve either three rescatterings in one nucleus
and one rescattering in the other nucleus or two rescatterings in each
of the nuclei.


\section{Asymmetric Collisions of Shock Waves in AdS$_5$}
\label{pA}


\subsection{Derivation of the Equations}

\begin{figure}[h]
\begin{center}
\epsfxsize=7cm
\leavevmode
\hbox{\epsffile{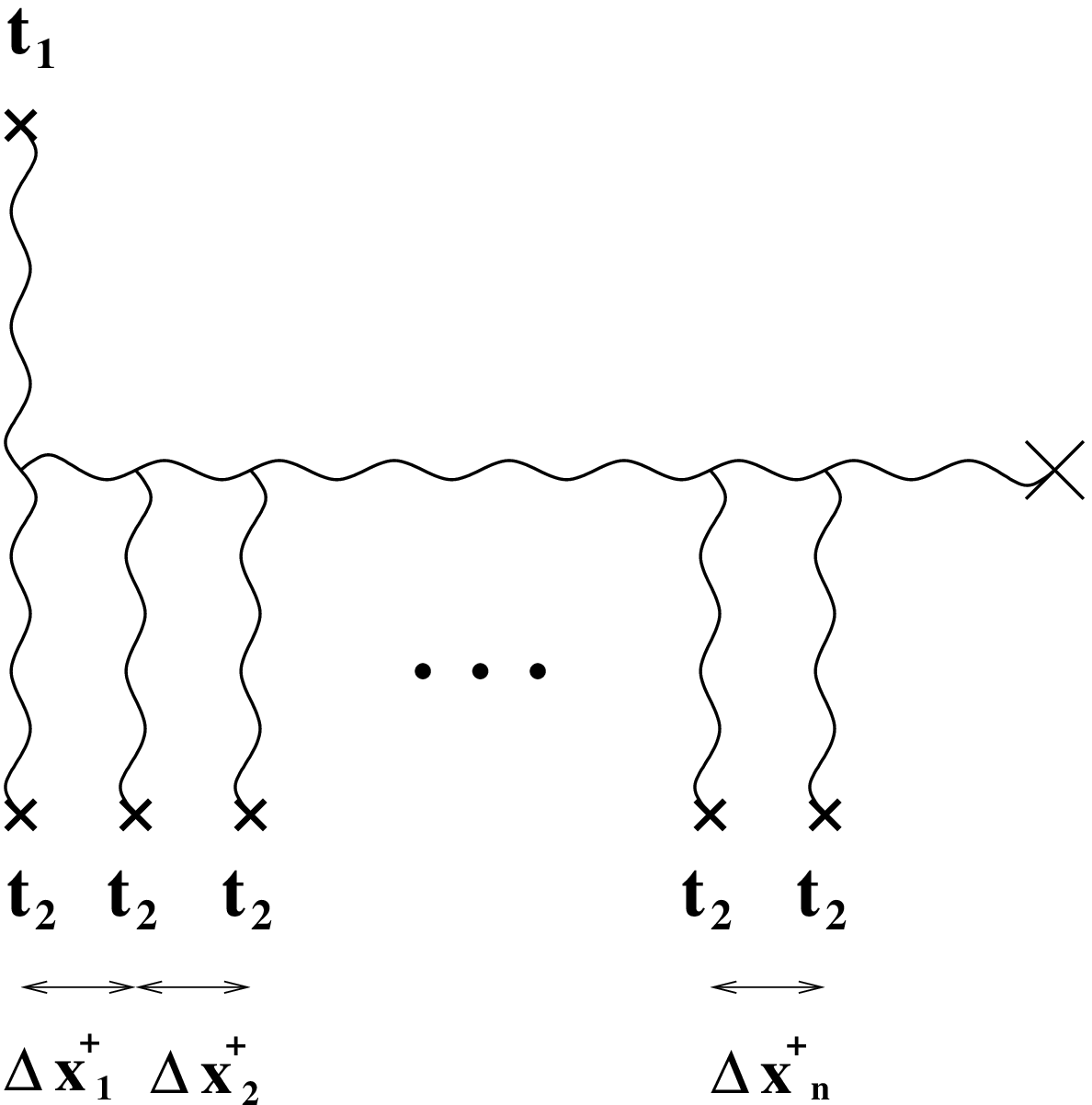}}
\end{center}
\caption{A diagram contributing to the metric of an asymmetric 
collision of two shock waves as considered in Section \protect\ref{pA}.}
\label{pAfig}
\end{figure}

We now want to find the solution of the proton-nucleus scattering
problem at strong coupling. In other words we want to resum all-order
graviton exchanges with one shock wave while keeping only terms with a
single graviton exchange with the second nucleus. That is, we want to
resum all powers of, say, $t_2$, while keeping only the leading power
of $t_1$. An example of a typical diagram which is resummed this way
is shown in \fig{pAfig}.

To resum the diagrams of the type shown in \fig{pAfig} let us first
construct the corresponding Einstein equations describing this
classical graviton field. We start by writing the metric, which is
just the same as given in \eq{2nuc_gen}, but without capitalizing the
unknown functions, to distinguish from the case of the full
nucleus-nucleus collisions:
\begin{align}\label{pA_gen}
  ds^2 \, = \, \frac{L^2}{z^2} \, \bigg\{ -\left[ 2 + g (x^+, x^-, z)
  \right] \, dx^+ \, dx^- + \left[ t_1 (x^-) \, z^4 + f (x^+, x^-, z)
  \right] \, d x^{- \, 2} \notag \\ + \left[ t_2 (x^+) \, z^4 +
    {\tilde f} (x^+, x^-, z) \right] \, d x^{+ \, 2} + \left[ 1 + h
    (x^+, x^-, z) \right] \, d x_\perp^2 + d z^2 \bigg\}.
\end{align}
We now want to plug the metric (\ref{pA_gen}) into the Einstein
equations (\ref{ein}) and linearize it in $t_1$. In doing so we have
to remember that, as $f$, $\tilde f$, $g$, and $h$ should have only
one factor of $t_1$ in them, one has $f, {\tilde f}, g, h \sim t_1$.
Thus one has to linearize Einstein equations in $t_1$ and in $f$,
$\tilde f$, $g$ and $h$. The relevant equations are
\begin{subequations}\label{ein_pA}
\begin{align}
  (\bot \bot) \hspace*{0.15in} & - 4 \, z^3 \, t_2 \, f - 8 \, z^7 \,
  t_1 \, t_2 - z^4 \, t_2 \, f_z \, + g_z + 5 \, h_z - z \, h_{z\, z}
  + z^5 \, t_2 \, h_{x^- \, x^-} + 2 \, z \, h_{x^+ \, x^-}   = 0 \label{pp} \\
  (zz) \hspace*{0.15in} & 8 \, z^3 \, t_2 \, f + 32 \, z^7 \, t_1 \,
  t_2 + 3 \, z^4 \, t_2 \, f_z + g_z + 2 \, h_z + z^5 \, t_2 \, f_{z\,
    z} - z \, g_{z\, z} - 2 \, z \, h_{z\, z} = 0 \label{zz} \\
  (-z) \hspace*{0.15in} & z^7 \, t'_1 \, t_2 \, + z^3 \, t_2 \,
  f_{x^-} - \frac{1}{4} \, g_{x^- \, z} - h_{x^- \, z} - \frac{1}{2}
  \, f_{x^+ \, z} =0, \label{-z}
\end{align}
\end{subequations}
where we suppressed the arguments of all functions and, as usual, the
subscripts $z$, $x^+$ and $x^-$ indicate partial derivatives with
respect to these variables. Again, the prime in $t'_1 (x^-)$ (and in
$t_2' (x^+)$ below) indicates a derivative with respect to the only
argument of the function. Other components of Einstein equations are
not needed, as Eqs.  (\ref{ein_pA}) contain enough information to find
$f$, $g$ and $h$. In fact we will need to know only $h$: as was shown
in Eqs.  (\ref{prod_Tmn2}) we can reconstruct the whole
energy-momentum tensor of the produced matter from it.

Solving \eq{pp} for $g_z$ and using the result to eliminate $g$ from
\eq{-z} yields
\begin{align}\label{*}
  \frac{1}{4} \, h_z - \frac{1}{2} \, \frac{\partial_+}{\partial_-} \,
  f_z - \frac{1}{4} \, z \, \left[ h_{z\, z} + z^3 \, t_2 \, (4 \, z^3
    \, t_1 + f_z - z \, h_{x^- \, x^-}) - 2 \, h_{x^+ \, x^-} \right]
  = 0.
\end{align}
Eliminating $g_z$ from \eq{zz} and solving the resulting equation for
$f_z$ we get
\begin{align}\label{fz}
  f_z \, = \, \frac{1}{4 \, z^4 \, t_2} \, \left[ - 16 \, z^7 \, t_1
    \, t_2 - 3 \, h_z + 3 \, z \, h_{z \, z} - z^2 \, h_{z \, z \, z}
    + 4 \, z^5 \, t_2 \, h_{x^- \, x^-} + z^6 \, t_2 \, h_{x^- \, x^-
      \, z} + 2 \, z^2 \, h_{x^+ \, x^- \, z} \right].
\end{align}
Applying an operator $\partial_- / \partial_+$ to \eq{*} and
substituting $f_z$ from \eq{fz} into it we obtain the following
equation for $h$
\begin{align}\label{h_pA}
  - 3 \, h_z + 3 \, z \, h_{z \, z} & - z^2 \, h_{z \, z \, z} + 2 \,
  z^2 \, h_{x^+ \, x^- \, z} = 16 \, z^7 \, t_1 \, t_2 \notag \\ & +
  z^4 \, t_2 \, \frac{\partial_-}{\partial_+} \, \left[ \frac{7}{2} \,
    h_z - \frac{7}{2} \, z \, h_{z \, z} + \frac{1}{2} \, z^2 \, h_{z
      \, z \, z} - 2 \, z^2 \, h_{x^+ \, x^- \, z} - \frac{1}{2} \,
    z^6 \, t_2 \, h_{x^- \, x^- \, z} \right].
\end{align}
Note that the first line of \eq{h_pA} is identical to the LO equation
(4.10) in \cite{Albacete:2008vs}. Higher order powers of $t_2$ come in
through the second line of \eq{h_pA}. 

\eq{h_pA} is the equation we need to solve. We slightly simplify it
by writing it as
\begin{align}\label{hmain}
  & z^2 \, \partial_z \, \left[ \frac{3}{z} \, h_z - h_{z \, z} + 2 \,
    h_{x^+ \, x^-} \right] \, = \, 16 \, z^7 \, t_1 \, t_2 \notag \\ & + z^4 \,
  t_2 \, \frac{\partial_-}{\partial_+} \, \left\{ z^2 \, \partial_z \,
    \left[ - \frac{7}{2} \, \frac{1}{z} \, h_z + \frac{1}{2} \, h_{z
        \, z} - 2 \, h_{x^+ \, x^-} \right] - \frac{1}{2} \, z^4 \,
    t_2 \, \partial_-^2 \, z^2 \, \partial_z h \right\}.
\end{align}
Below we will use \eq{hmain} to evaluate the diagram in \fig{pAfig} in
the eikonal approximation, which we will define in the next
Subsection.


\subsection{Green Function and the Eikonal Approximation}

To construct the solution of \eq{hmain} we will need to construct the
retarded Green function of the operator on its left hand side. As
inverting $z^2 \, \partial_z$ is trivial, we will need the function $G
(x^+, x^-, z; x'^+, x'^-, z')$ such that
\begin{align}\label{G1}
  \left[ \frac{3}{z} \, \partial_z - \partial_z^2 + 2 \, \partial_+ \,
    \partial_- \right] \, G (x^+, x^-, z; x'^+, x'^-, z') \, = \,
  \delta (x^+ - x'^+) \, \delta (x^- - x'^-) \, \delta (z-z').
\end{align}
This is a bulk-to-bulk scalar field propagator, which has previously
been found in \cite{Danielsson:1998wt}. For completeness of the
presentation let us briefly outline the construction of $G (x^+,
x^-, z; x'^+, x'^-, z')$. 

Fourier-transforming \eq{G1} into light-cone momentum space (i.e.,
going from $x^+$ and $x^-$ coordinates to their conjugates $k^+$,
$k^-$ but keeping the coordinate $z$) and dropping the delta-function
on the right one can see that the solution of the resulting equation
is simply $z^2 \, J_2 (z \sqrt{2 \, k^+ \, k^-})$. Using these Bessel
function and going back to the $x^0$, $x^3$ coordinates instead of
$x^+$, $x^-$ we write for the retarded Green function
\begin{align}\label{G2}
  G (x^0, x^3, z; x'^0, x'^3, z') \, = \, \frac{\theta (x^0 - x'^0)}{2
    \, \pi} \, \int\limits_0^\infty d m \,
  \int\limits_{-\infty}^\infty dk \, \frac{\sin \left[ (x^0 - x'^0)
      \sqrt{m^2 + k^2} \right]}{ \sqrt{m^2 + k^2}} \, e^{i \, k \,
    (x^3 - x'^3)} \notag \\ \times \, m \, z^2 \, J_2 (m \, z) \,
  \frac{1}{z'} \, J_2 (m \, z').
\end{align}
The integral over the momentum variable $k$ can be performed yielding
\begin{align}\label{G3}
  G (x^+, x^-, z; x'^+, x'^-, z') \, & = \, \frac{1}{2} \, \theta (x^+
  - x'^+) \, \theta (x^- - x'^-) \, \frac{z^2}{z'} \,
  \int\limits_0^\infty d m \notag \\ & \times \, m \, J_0\left( m \,
    \sqrt{2 \, (x^+ - x'^+) \, (x^- - x'^-)} \right) \, J_2 (m \, z)
  \, J_2 (m \, z').
\end{align}
\eq{G3} can be further simplified by integration over $m$, which gives
\begin{align}\label{G4}
  G (x^+, x^-, z; x'^+, x'^-, z') \, & = \, \frac{1}{2 \, \pi} \,
  \theta (x^+ - x'^+) \, \theta (x^- - x'^-) \, \theta (s) \, \theta
  (2-s) \, \frac{z}{z'^2} \, \frac{1 + 2 \, s \, (s-2)}{\sqrt{s \,
      (2-s)}}
\end{align}
with
\begin{align}
  s \equiv \frac{2 \, (x^+ - x'^+) \, (x^- - x'^-) - (z-z')^2}{2 \, z
    \, z'}.
\end{align}
However \eq{G3} is really all we need for the calculations to follow.

\FIGURE[th]{\includegraphics[width=7cm]{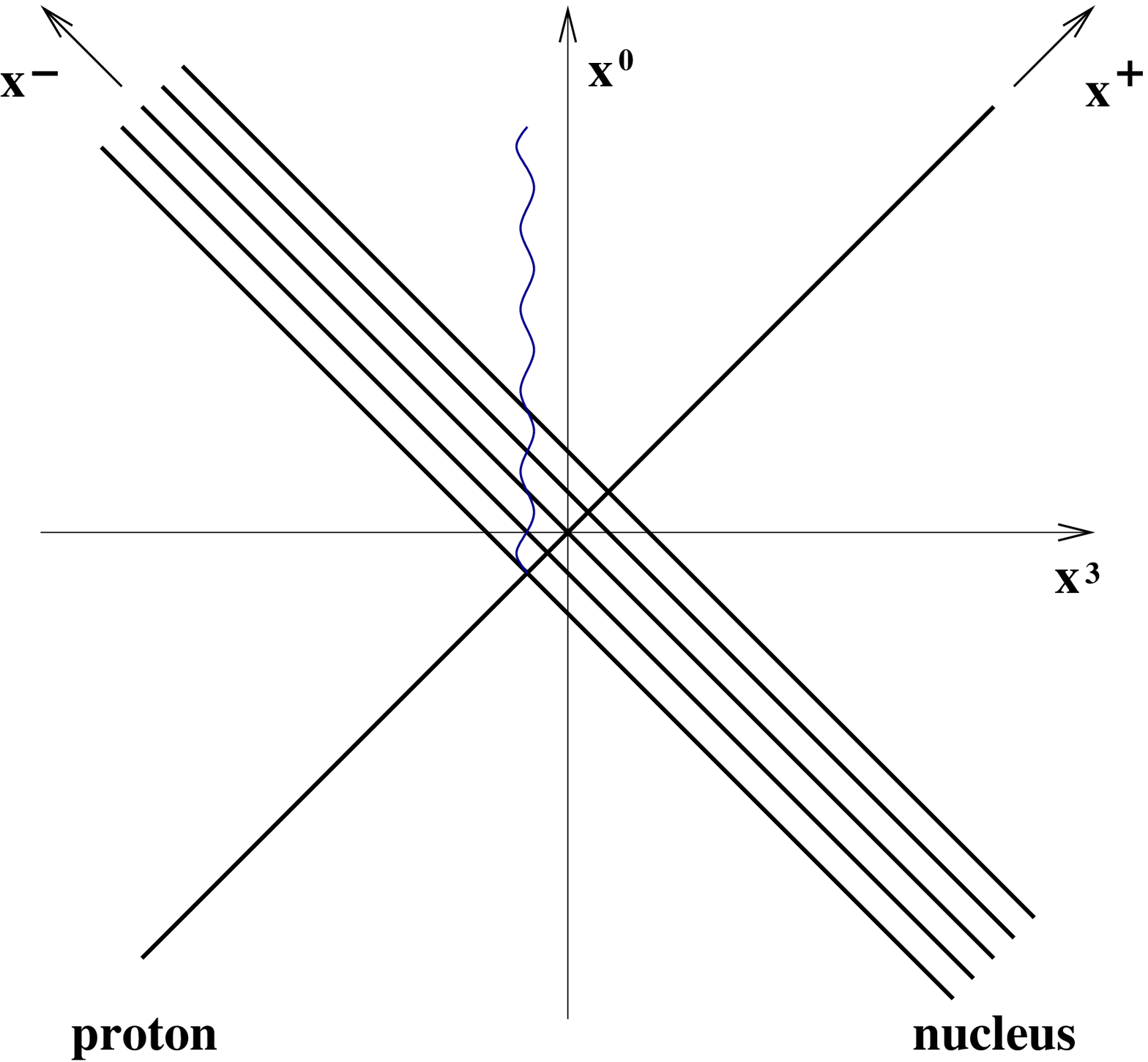}
  \caption{The space-time structure of the graviton emission in a 
    proton--nucleus collision. The graviton is denoted by the wavy
    line. After being produced the graviton rescatters in the nucleus
    and the propagates freely in the forward light-cone. }
  \label{pAsp}
}

Eqs. (\ref{G3}) or (\ref{G4}) give us the propagator of the
gravitons in the $s$-channel of \fig{pAfig}. These expressions allow
us to construct the eikonal approximation for the graviton production
in asymmetric shock wave collisions. The space-time structure of
graviton production in such collisions in shown in \fig{pAsp}. It
illustrates the diagram in \fig{pAfig}: first the graviton (the wavy
line) is produced in a collision of the proton shock wave and some
elements of the nucleus shock wave (a nucleon in the nucleus). This
generates the LO factor of $t_1 \, t_2$. Subsequently the graviton
rescatters in the nucleus shock wave with each rescattering bringing
in a factor of $t_2$. After the graviton leaves the shock wave it
simply propagates freely. Indeed the transverse dimensions $x^1, x^2$
and the 5th dimension in AdS$_5$ are implied but not shown in
\fig{pAsp}.

Most importantly, the propagation of the graviton between two
successive rescatterings in the nucleus shock wave happens over a very
short interval in the light-cone ``plus'' direction. Namely the
intervals $\Delta x^+_i$'s between the rescatterings in \fig{pAfig}
are Lorentz-contracted and are all of the order $\Delta x^+_i \sim
1/p_2^-$. As $p_2^-$ (along with the comparable scale $p_1^+$) is the
largest momentum scale in the problem we conclude that $\Delta
x^+_i$'s are the shortest distance scales in the problem, i.e., they
are {\sl very small} compared to any other distance scale. This is
illustrated in \fig{pAsp}, which depicts the propagation of the
graviton through the highly Lorentz-contracted nucleus. Therefore we
can approximate the full $s$-channel graviton propagator by its
short-$x^+$-interval version. We will call such approximation an {\sl
  eikonal approximation} in analogy with the terminology used in high
energy scattering in four dimensions.

Putting $x^+ \approx x'^+$ in \eq{G3} we can put $J_0 (0) =1$ which
yields the Green function in the eikonal approximation
\begin{align}\label{Geik1}
  G_{eik} (x^+, x^-, z; x'^+ \approx x^+ , x'^-, z') \, & = \,
  \frac{1}{2} \, \theta (x^+ - x'^+) \, \theta (x^- - x'^-) \,
  \frac{z^2}{z'} \, \int\limits_0^\infty d m \, m \, \, J_2 (m \, z)
  \, J_2 (m \, z') \notag \\ & = \, \frac{1}{2} \, \theta (x^+ - x'^+)
  \, \theta (x^- - x'^-) \, \delta (z - z').
\end{align}
For inhomogeneous equations like (\ref{hmain}), or like the following
equation
\begin{align}\label{hR}
  \frac{3}{z} \, h_z - h_{z \, z} + 2 \, h_{x^+ \, x^-} \, = \, R
  (x^+, x^-, z),
\end{align}
in the solution, the Green function acts on some function $R (x^+,
x^-, z)$ on the right hand side, such that
\begin{align}\label{Gsol}
  h (x^+, x^-, z) \, = \, \int\limits_{-\infty}^\infty d x'^+ \,
  \int\limits_{-\infty}^\infty d x'^- \, \int\limits_0^\infty d z' \,
  G (x^+, x^-, z; x'^+, x'^-, z') \, R (x'^+, x'^-, z').
\end{align}
Using the eikonal Green function (\ref{Geik1}) in \eq{Gsol} yields
\begin{align}\label{heik1}
  h_{eik} (x^+, x^-, z) \, = \, \frac{1}{2 \, \partial_+ \,
    \partial_-} \, R (x^+, x^-, z)
\end{align}
with the inverse derivatives defined in \eq{ints}.

Going from Eq. (\ref{hR}) to Eq. (\ref{heik1}) clarifies the procedure
for the eikonal approximation: simply neglecting all $z$-derivatives
on the left hand side of \eq{hR} compared to $\partial_+$ we obtain
\eq{heik1}. Indeed $\partial_+ \sim 1/\Delta x^+ \sim p^-_2$ if the
propagator in question spans a short interval $\Delta x^+$. Hence the
main rule of the eikonal approximation is that $\partial_+$ is much
larger than any other derivative in the problem.  The short interval
scalar field bulk-to-bulk propagator is
\begin{align}\label{Geik2}
  \int\limits_{-\infty}^\infty d x'^+ \, \int\limits_{-\infty}^\infty
  d x'^- \, \int\limits_0^\infty d z' \, G_{eik} (x^+, x^-, z; x'^+,
  x'^-, z') \, \left[ \ldots \right] \, = \, \frac{1}{2 \, \partial_+
    \, \partial_-} \, \left[ \ldots \right].
\end{align}

One has to keep in mind that the eikonal approximation should be
applied to short-lived propagators only. That is we can not just take
\eq{hmain} and drop all terms not containing $\partial_+$. As can be
seen from Figs. \ref{pAfig} and \ref{pAsp}, the graviton propagator
after the interaction with the nucleus is not limited to any short
interval in any direction. That is, we have to use the full propagator
(\ref{G3}) for that line. Note that we are not calculating the
graviton production amplitude: we are calculating the graviton field.
Hence in the diagram in \fig{pAfig} the outgoing graviton propagator
is off-mass shell, and is {\sl not} on mass shell, as it would have
been for the production amplitude. (See e.g. \cite{Kovchegov:1997pc}
for an example of constructing Feynman diagrams corresponding to
classical fields.)

The graviton propagator in \eq{Geik2} does not take into account
rescatterings and only describes free propagation for a graviton over
a short time interval. The eikonal approximation should also be
applied to the multi-graviton vertices in \fig{pAfig}. To facilitate
the application of the eikonal approximation let us recast \eq{hmain}
in a slightly different form.  Defining
\begin{align}
  {\tilde h} \, = \, z^2 \, \partial_z \, h
\end{align}
we rewrite \eq{hmain} as
\begin{align}\label{htmain}
  \left[ {\hat D}_1 + 2 \, \partial_+ \, \partial_- \right] \, {\tilde
    h} \, = \, 16 \, z^7 \, t_1 \, t_2 + z^4 \, t_2 \,
  \frac{\partial_-}{\partial_+} \, \left\{ \left[ {\hat D}_2 - 2 \,
      \partial_+ \, \partial_- \right] {\tilde h} - \frac{1}{2} \, z^4
    \, t_2 \, \partial_-^2 \, {\tilde h} \right\}
\end{align}
where we have defined differential operators
\begin{align}\label{D1def}
  {\hat D}_1 \, = \, z^2 \, \partial_z \, \left[ \frac{5}{z^3} -
    \frac{1}{z^2} \, \partial_z \right]
\end{align}
and
\begin{align}\label{D2def}
  {\hat D}_2 \, = \, z^2 \, \partial_z \, \left[ - \frac{9}{2} \,
    \frac{1}{z^3} + \frac{1}{2} \, \frac{1}{z^2} \, \partial_z
  \right].
\end{align}
Defining the truncated amplitude 
\begin{align}
  {\bar h} \, = \, \left[ {\hat D}_1 + 2 \, \partial_+ \, \partial_-
  \right] \, {\tilde h}
\end{align}
allows us to write \eq{htmain} as
\begin{align}\label{hbar1}
  {\bar h} \, = \, 16 \, z^7 \, t_1 \, t_2 + z^4 \, t_2 \,
  \frac{\partial_-}{\partial_+} \, \left[ \left( {\hat D}_2 - 2 \,
      \partial_+ \, \partial_- \right) \, \left( {\hat D}_1 + 2 \,
      \partial_+ \, \partial_- \right)^{-1} - \frac{1}{2} \, z^4 \,
    t_2 \, \partial_-^2 \, \, \left( {\hat D}_1 + 2 \, \partial_+ \,
      \partial_- \right)^{-1} \right] \, {\bar h}.
\end{align}
The solution of \eq{hbar1} is
\begin{align}\label{hbar3}
  {\bar h} \, = \, & \left[ 1 + z^4 \, t_2 \,
    \frac{\partial_-}{\partial_+} \, \left(1 - \frac{{\hat D}_2}{ 2 \,
        \partial_+ \, \partial_-} \right) \, \left( 1 + \frac{{\hat
          D}_1}{2 \, \partial_+ \, \partial_-} \right)^{-1} +
    \frac{1}{4} \, \left( z^4 \, t_2 \, \frac{\partial_-}{\partial_+}
    \right)^2 \, \left( 1 + \frac{{\hat D}_1}{2 \, \partial_+ \,
        \partial_-} \right)^{-1} \right]^{-1} \notag \\ & \times \, 16
  \, z^7 \, t_1 \, t_2.
\end{align}
As $t_2$ and $1/\partial_+$ do not commute, here and throughout the
paper we have
\begin{align}
  \left( t_2 \, \frac{1}{\partial_+} \right)^2 \, \ldots \, = \, t_2
  \, \frac{1}{\partial_+} \left( t_2 \, \frac{1}{\partial_+} \ldots
  \right),
\end{align}
that is, each $1/\partial_+$ operator acts on {\sl everything} to its
right.

A simple algebra gives
\begin{align}\label{hhbar}
  h = \left[ \frac{3}{z} \, \partial_z - \partial_z^2 + 2 \,
    \partial_+ \, \partial_- \right]^{-1} \, \frac{1}{\partial_z} \,
  \left( \frac{\bar h}{z^2} \right)
\end{align}
with
\begin{align}
  \frac{1}{\partial_z} \, [\ldots] (z) \, = \, \int\limits_0^z dz' \,
  [\ldots] (z') \, .
\end{align}
Therefore $h$ and $\bar h$ are related to each other with the help of
the Green function (\ref{G3}). Therefore $\bar h$ is really the part
of the amplitude in \fig{pAfig} without the last $s$-channel gluon
propagator, i.e., $\bar h$ is the truncated amplitude. As all
$s$-channel graviton propagators in the truncated amplitude $\bar h$
are short-lived, we can apply the eikonal approximation to the
equation (\ref{hbar3}) for $\bar h$. In fact \eq{hbar3} is already
cast in the form designed to simplify the expansion in inverse powers
of $\partial_+$. The eikonal $\bar h$ we obtain this way can be used
in \eq{hhbar} with the full Green function (\ref{G3}) to recover $h
(x^+, x^-, z)$. As \eq{hbar3} appears to be prohibitively complicated
to evaluate analytically, the eikonal approximation appears to be the
only way to proceed. In fact, as we will shortly see, for the
delta-function shock waves it yields the exact solution for the metric
generated in the asymmetric (proton-nucleus) collision of two shock
waves.


\subsection{Delta-Function Shock Waves}
\label{delta_sec}


\subsubsection{Energy-Momentum Tensor of the Produced Medium}

Let us again consider a collision of two physical delta-function shock
waves with $t_1$ and $t_2$ given by \eq{deltas}. We will also keep the
smeared shock waves in \eq{thetas} in mind. 

First let us apply the eikonal approximation to \eq{hbar3} without
substituting the explicit expressions for $t_1$ and $t_2$ from
\eq{deltas}. As we argued above, in the eikonal approximation the
derivative $\partial_+$ is the largest momentum scale in the problem.
Hence in \eq{hbar3} we have
\begin{align}\label{Deik}
  \frac{{\hat D}_1}{2 \, \partial_+ \, \partial_-} \, \ll 1, \ \ \ 
  \frac{{\hat D}_2}{2 \, \partial_+ \, \partial_-} \, \ll 1.
\end{align}
After neglecting those terms \eq{hbar3} yields
\begin{align}\label{hbar4}
  {\bar h}_{eik} \, = \, \left[ \left( 1 + \frac{1}{2} \, z^4 \, t_2
      \, \frac{\partial_-}{\partial_+} \right)^2 \right]^{-1} \, 16 \,
  z^7 \, t_1 \, t_2.
\end{align}
To evaluate \eq{hbar4} we expand it in a series
\begin{align}\label{hbar5}
  {\bar h}_{eik} \, = \, \sum\limits_{n=0}^\infty \, (n+1) \, \left( -
    \frac{1}{2} \, z^4 \, t_2 \, \frac{\partial_-}{\partial_+}
  \right)^n \, 16 \, z^7 \, t_1 \, t_2.
\end{align}
Using \eq{hbar5} in \eq{hhbar} along with \eq{G3} we write
\begin{align}\label{heik2}
  h_{eik} \, = \, \int\limits_{-\infty}^\infty d x'^+ \,
  \int\limits_{-\infty}^\infty d x'^- \, \int\limits_0^\infty d z' \,
  G (x^+, x^-, z; x'^+, x'^-, z') \, \frac{1}{\partial_{z'}} \, \left(
    \frac{{\bar h}_{eik}}{z'^2} \right) \notag \\
  = \, \int\limits_{-\infty}^{x^+} d x'^+ \,
  \int\limits_{-\infty}^{x^-} d x'^- \, \int\limits_0^\infty d z' \,
  \frac{1}{2} \, \frac{z^2}{z'} \, \int\limits_0^\infty d m \, m \,
  J_0\left( m \, \sqrt{2 \, (x^+ - x'^+) \, (x^- - x'^-)} \right) \,
  J_2 (m \, z) \, J_2 (m \, z')  \notag \\
  \times \, \sum\limits_{n=0}^\infty \, \frac{n+1}{2 n + 3} \, \left(
    - \frac{1}{2} \, t_2 (x'^+) \, \frac{\partial'_-}{\partial'_+}
  \right)^n \, 8 \, (z')^{4 n + 6} \, t_1 (x'^-) \, t_2 (x'^+) .
\end{align}
Here $\partial'_\pm = \partial/\partial x'^\pm$. The expression
(\ref{heik2}) is still rather difficult to evaluate. However, as we
are primarily interested in the dynamics of the gauge theory, we only
need the $z^4$ term in this expression to obtain the transverse
pressure of the produced medium using Eqs. (\ref{prod_Tmn2}) and
(\ref{pdef}). As the series expansion of the Bessel functions
converges everywhere, we expand
\begin{align}\label{Bess}
  J_2 (m \, z) = \frac{1}{8} \, m^2 \, z^2 + o(z^4)
\end{align}
in \eq{heik2} and integrate over $z'$ and $m$ obtaining the transverse
pressure
\begin{align}\label{peik1}
  p \, = \, \frac{N_c^2}{2 \, \pi^2} \, 8 \, \sum\limits_{n=0}^\infty
  \, (n+1)^2 \, (-2)^n \, \int\limits_{-\infty}^{x^+} d x'^+ \,
  \int\limits_{-\infty}^{x^-} d x'^- \, (x^+ - x'^+)^{1 + 2 n} \, (x^-
  - x'^-)^{1 + 2 n} \notag \\ \times \, \left[ \partial'^{\, n}_- \,
    t_1 (x'^-) \right] \, \left( t_2 (x'^+) \, \frac{1}{\partial'_+}
  \right)^n \, t_2 (x'^+).
\end{align}
Using integration by parts in the integral over $x'^-$ in \eq{peik1}
and remembering that $t_1$ is a localized function of $x^-$ yields the
final expression for the transverse pressure
\begin{align}\label{peik}
  p \, = \, \frac{N_c^2}{2 \, \pi^2} \, 8 \, \sum\limits_{n=0}^\infty
  \, (-2)^n \, (n+1) \, \frac{(2 \, n + 1)!}{n!} \,
  \int\limits_{-\infty}^{x^-} d x'^- \, (x^- - x'^-)^{1 + n} \, t_1
  (x'^-) \notag \\ \times \, \int\limits_{-\infty}^{x^+} d x'^+ \,
  (x^+ - x'^+)^{1 + 2 n} \, \left( t_2 (x'^+) \, \frac{1}{\partial'_+}
  \right)^n \, t_2 (x'^+).
\end{align}
\eq{peik} is one of the main results of this Section. It is the
simplest expression for $p$ we could find without using an explicit
form for the functions $t_1$ and $t_2$.

As with the NLO calculations of Sect. \ref{NLOdeltas}, substituting
$t_1$ and $t_2$ from \eq{deltas} into \eq{peik} would generate terms
like $\delta (x^+) \, \theta (x^+)$, evaluation of which is ambiguous.
To avoid this ambiguity we use the smeared $t_1$ and $t_2$ from
\eq{thetas}. For $x^-\gg a_1$ and $x^+ \gg a_2$ we have
\begin{align}\label{t1d}
  \int\limits_{-\infty}^{x^-} d x'^- \, (x^- - x'^-)^{1 + n} \, t_1
  (x'^-) \, \approx \, \mu_1 \, (x^-)^{1+n} \, \theta (x^-)
\end{align}
and
\begin{align}\label{t2d}
  \int\limits_{-\infty}^{x^+} d x'^+ \, (x^+ - x'^+)^{1 + 2 n} \,
  \left( t_2 (x'^+) \, \frac{1}{\partial'_+} \right)^n \, t_2 (x'^+)
  \, \approx \, \frac{1}{(n+1)!} \, \mu_2^{n+1} \, (x^+)^{1 + 2 \, n}
  \, \theta (x^+).
\end{align}
Using Eqs. (\ref{t1d}) and (\ref{t2d}) in \eq{peik} and summing the
series over $n$ yields
\begin{align}\label{pd}
  p \, = \, \frac{N_c^2}{2 \, \pi^2} \, \frac{8 \, \mu_1 \, \mu_2 \,
    x^+ \, x^- \, \theta (x^+) \, \theta (x^-)}{\left[ 1 + 8 \, \mu_2
      \, (x^+)^2 \, x^- \right]^{3/2}}.
\end{align}
This is another main result of this Section: \eq{pd} gives us the
transverse pressure of the medium produced in the collision of a
proton and a nucleus at strong coupling. It is valid at $x^- \gg a_1$
and $x^+ \gg a_2$: these conditions are automatically satisfied if the
sources are exact delta-functions of \eq{deltas}. Hence for the
delta-function sources (\ref{deltas}) \eq{pd} provides us with the
exact solution of the problem! As a cross-check one can see that
expanding \eq{pd} in a series in $\mu_2$ to NLO yields \eq{pNLO}.

\eq{pd} allows us to explicitly specify the limits of our
approximation. Namely, we resum all powers of rescattering in the
nucleus, which, for delta-function shock waves translate into powers
of $\mu_2 \, (x^+)^2 \, x^-$. At the same time we neglect higher
rescatterings in the proton, which, by analogy, would bring in powers
of $\mu_1 \, (x^-)^2 \, x^+$. Hence the applicability region of
\eq{pd} is defined by
\begin{align}\label{bounds}
  \mu_1 \, (x^-)^2 \, x^+ \, \ll \, 1, \ \ \ \mu_2 \, (x^+)^2 \, x^-
  \sim 1.
\end{align}
(Indeed for small $\mu_2 \, (x^+)^2 \, x^-$ \eq{pd} applies too.) For
non-delta function shock waves like those given in \eq{thetas} one
also has to keep the limit (\ref{bounds2}) in mind while studying the
applicability region of \eq{pd}.

Using \eq{pd} along with \eq{prod_Tmn2} we can find all other non-zero
components of the energy-momentum tensor of the produced medium:
\begin{subequations}\label{Tmnd}
\begin{align}
  \langle T^{++}\rangle \, & = \, - \frac{N_c^2}{2 \, \pi^2} \,
  \frac{4 \, \mu_1 \, \mu_2 \, (x^+)^2 \, \theta (x^+) \, \theta
    (x^-)}{\left[ 1 + 8 \, \mu_2 \, (x^+)^2 \, x^- \right]^{3/2}}, \label{T++} \\
  \langle T^{--}\rangle \, & = \, \frac{N_c^2}{2 \, \pi^2} \, \theta
  (x^+) \, \theta (x^-) \, \frac{\mu_1}{2 \, \mu_2 \, (x^+)^4} \notag
  \\ & \times \, \frac{3 - 3 \, \sqrt{1 + 8 \, \mu_2 \, (x^+)^2 \,
      x^-} + 4 \, \mu_2 \, (x^+)^2 \, x^- \, \left( 9 + 16 \, \mu_2 \,
      (x^+)^2 \, x^- - 6 \, \sqrt{1 + 8 \, \mu_2 \, (x^+)^2 \, x^-}
    \right) }{\left[ 1 + 8 \, \mu_2 \, (x^+)^2 \, x^- \right]^{3/2}}, \\
  \langle T^{+-}\rangle \, & = \, \frac{N_c^2}{2 \, \pi^2} \, \frac{8
    \, \mu_1 \, \mu_2 \, x^+ \, x^- \, \theta (x^+) \, \theta
    (x^-)}{\left[ 1 + 8 \, \mu_2 \, (x^+)^2 \, x^- \right]^{3/2}}, \\
  \langle T^{\, i j}\rangle \, & = \, \delta^{ij} \, \frac{N_c^2}{2
    \,\pi^2} \, \frac{8 \, \mu_1 \, \mu_2 \, x^+ \, x^- \, \theta
    (x^+) \, \theta (x^-)}{\left[ 1 + 8 \, \mu_2 \, (x^+)^2 \, x^-
    \right]^{3/2}}.
\end{align}
\end{subequations}
Provided the complexity of the problem at hand, the resulting formulas
(\ref{Tmnd}) for the energy-momentum tensor are remarkably simple!

Now we can ask a question: what kind of medium is produced in these
strongly coupled proton-nucleus collisions? Is it described by ideal
hydrodynamics, just like Bjorken hydrodynamics was obtained in
\cite{Janik:2005zt}? In our case the produced matter distribution is
obviously rapidity-dependent, so it is slightly more tricky to check
whether Eqs. (\ref{Tmnd}) constitute an ideal hydrodynamics, i.e.,
whether it can be written as
\begin{align}\label{hydro}
  T^{\mu\nu} \, = \, (\epsilon + p) \, u^\mu \, u^\nu - p \,
  \eta^{\mu\nu}
\end{align}
with the positive energy density $\epsilon$ and pressure $p$.
$\eta^{\mu\nu}$ is the metric of the four-dimensional Minkowski
space-time and $u^\mu$ is the fluid 4-velocity.

For the particular case at hand it is easy to see that the
energy-momentum tensor in \eq{Tmnd} can not be cast in the ideal
hydrodynamics form of (\ref{hydro}). In the case of ideal
hydrodynamics one has
\begin{align}
  T^{++} \, = \, (\epsilon + p) \, (u^+)^2 \, > \, 0.
\end{align}
At the same time $\langle T^{++}\rangle$ in \eq{T++} is {\sl negative
  definite}. Therefore the ideal hydrodynamic description is not
achieved in the proton-nucleus collisions. We believe this result is
due to limitations of this proton-nucleus approximation. Any strongly
coupled medium at asymptotically late times is almost certainly bound
to thermalize. Our conclusion is then that
thermalization/isotropization of the medium does not happen in the
space-time region defined by the bounds in \eq{bounds}. What we found
in \eq{Tmnd} is a medium at some intermediate stage, presumably on its
way to thermalization at a later time.  It is likely that one needs to
solve the full nucleus-nucleus scattering problem to all orders, as
shown in \fig{AA}, to obtain a medium described by ideal
hydrodynamics.


\subsubsection{Proton Stopping}\label{prot-stop}

In \cite{Albacete:2008vs} it was argued that the physical shock waves
given by Eq. (\ref{deltas}) or by \eq{thetas} come to a complete stop
shortly after the collision. The conclusion was based on the LO
calculation, which for the shock waves (\ref{thetas}) gave the
following $++$ component of the energy momentum tensor of a nucleus
(or its remnants) moving in the light-cone plus direction after the
collision
\begin{align}\label{stopLO}
  \langle T^{++} (x^+ \gg a, x^- = a/2) \rangle \, = \,
  \frac{N_c^2}{2 \, \pi^2} \, \frac{\mu}{a} \, \left[ 1 - 2 \, \mu \,
    x^{+\, 2} \, a \right].
\end{align}
In arriving at \eq{stopLO} in \cite{Albacete:2008vs} we for simplicity
put $\mu_1 = \mu_2 = \mu$ and $a_1 = a_2 = a$. \eq{stopLO} allowed us
to conclude that as the light-cone time $x^+ \sim 1/\sqrt{\mu \, a}$
the $++$ component of the energy momentum tensor of the shock wave
would become zero, meaning that the shock wave stops propagating along
the light cone. Indeed, as we saw above (see \eq{bounds}), at the same
time as the stopping happens, i.e., when $ \mu \, x^{+\, 2} \, a \sim
1$, higher order graviton exchanges would become important.  With the
help of \eq{peik}
we can now explore whether multiple graviton exchanges with the
nucleus shock wave modify our conclusion about proton stopping reached
in \cite{Albacete:2008vs} at the LO level.




We start by evaluating \eq{peik} for $x^+ \gg a_2$, but with $0 < x^-
< a_1$. That way we follow the proton shock wave for some time after
the collision, which allows us to find the energy-momentum tensor of
the shock wave itself. As $x^+ \gg a_2$ still, \eq{t2d} remains
unchanged. We have to re-evaluate the left-hand-side of \eq{t1d} for
$0 < x^- < a_1$. This can be readily done yielding
\begin{align}\label{t1dp}
  \int\limits_{-\infty}^{x^-} d x'^- \, (x^- - x'^-)^{1 + n} \, t_1
  (x'^-) \, = \, \frac{\mu_1}{a_1} \, \frac{1}{n+2} \, (x^-)^{2+n}, \ 
  \ \ \mbox{for} \ \ \ 0 < x^- < a_1.
\end{align}
Using Eqs. (\ref{t2d}) and (\ref{t1dp}) in \eq{peik} and resumming the
series one obtains the transverse pressure inside the proton shock
wave, which, with the help of \eq{prod_Tmn2}, gives the following
expression for the $++$ component of the energy momentum tensor of the
produced matter
\begin{align}\label{prod}
  \langle T^{++}_{prod} \rangle \, = \, \frac{N_c^2}{2 \, \pi^2} \,
  \frac{\mu_1}{a_1} \, \left\{ -1 + \frac{1}{\sqrt{1 + 8 \, \mu_2 \,
        (x^+)^2 \, x^-}} \right\}, \ 
  \ \ \mbox{for} \ \ \ 0 < x^- < a_1.
\end{align}
Eqs. (\ref{t1}) and (\ref{thetas}) give the energy-momentum tensor of
the original incoming shock wave itself as
\begin{align}\label{orig}
  \langle T^{++}_{orig} \rangle \, = \, \frac{N_c^2}{2 \, \pi^2} \,
  \frac{\mu_1}{a_1}, \ \ \ \mbox{for} \ \ \ 0 < x^- < a_1.
\end{align}
Adding the energy-momentum tensors of the original shock wave and the
produced matter given in Eqs. (\ref{orig}) and (\ref{prod}) together
we obtain the total $++$ component of the energy-momentum tensor of
the proton shock wave
\begin{align}\label{stop}
  \langle T^{++}_{tot} \rangle \, = \, \langle T^{++}_{orig} \rangle +
  \langle T^{++}_{prod} \rangle \, = \, \frac{N_c^2}{2 \, \pi^2} \,
  \frac{\mu_1}{a_1} \, \frac{1}{\sqrt{1 + 8 \, \mu_2 \, (x^+)^2 \,
      x^-}},  \ \ \ \mbox{for} \ \ \ 0 < x^- < a_1.
\end{align}
Expanding \eq{stop} in the powers of $\mu_2$ at $x^- = a_1 /2$ would
yield \eq{stopLO}, providing an independent consistency check. 

\eq{stop} clearly demonstrates that the $++$ component of the
energy-momentum tensor of the proton shock wave is {\sl positive
  definite}. Notice that the LO solution (\ref{stopLO}) for $T^{++}$
becomes negative at large enough $x^+$. Inclusion of multiple
graviton exchanges fixes this problem. $T^{++}$ in \eq{stop} goes to
zero smoothly as $x^+$ grows large for any fixed $x^-$ in the $0 < x^-
< a_1$ range. Thus \eq{stop} explicitly demonstrates that
strong-coupling interactions of the proton shock wave with the nucleus
shock wave would stop the proton shock wave shortly after the
collision. For $x^- = a_1 /2$ the stopping happens at $x^+ \sim
1/\sqrt{\mu_2 \, a_1}$, in agreement with the arguments of
\cite{Albacete:2008vs}.


\subsection{Delta-Prime Shock Waves}


\subsubsection{Deltology}

The eikonal approximation used in Sect. \ref{delta_sec} reduces the
exact formula (\ref{hbar3}) to \eq{hbar4}. \eq{hbar4} resums the
powers of $t_2$ with only one factor of $1/\partial_+$ inserted
between each pair of $t_2$'s. It thus resums terms consisting of
sequences like
\begin{align}\label{Leik}
  t_2 \, \frac{1}{\partial_+} \, t_2 \, \frac{1}{\partial_+} \, t_2 \,
  \frac{1}{\partial_+} \, \ldots \, \frac{1}{\partial_+} \, t_2
\end{align}
(see also \eq{peik}).  This is indeed natural in the eikonal
approximation, as $\partial_+ \sim p_2^-$ is large and we want to
have as little powers of $1/\partial_+ \sim 1/p_2^-$ as possible in
each term.  \eq{hbar4} resums the absolute minimum number of the
powers of $1/\partial_+$.

An attentive reader might have noticed that the approximation of
\eq{hbar4} is insufficient for the delta-prime shock waves of
\eq{tzero}. Indeed performing the calculation in Appendix \ref{A} we
saw that leading powers of $p_2^-$ arose not only from the terms of
the type shown in \eq{Leik}, like we had in \eq{psi11}, but also from
terms with two powers of $1/\partial_+$ inserted between two $t_2$'s,
as can be seen from Eqs. (\ref{Bterm}) and (\ref{psi9}). Hence we need
to rethink our power counting if we want to resum the leading eikonal
terms for the delta-prime shock waves.

Let us start with delta-function shock waves with $t_2 \sim \delta
(x^+)$. In this case
\begin{align}\label{Leik1}
  t_2 \, \frac{1}{\partial_+} \, t_2 \, \sim \, \delta (x^+) \, \theta (x^+)
  \sim \, \delta (x^+).
\end{align}
Here we are being rather sloppy in treating $\delta (x^+) \, \theta
(x^+)$: of course the whole regularization introduced in \eq{thetas}
above was designed to obtain the correct values for $\theta (0)$ in
different situations. However, for the purposes of counting powers of
$p_2^-$ the exact value of $\theta (0)$ is not important as long as it
is a $p_2^-$-independent number. \eq{Leik1} demonstrates that for $t_2
\sim \delta (x^+)$ one has
\begin{align}\label{Leik2}
  t_2 \, \frac{1}{\partial_+} \, t_2 \, \frac{1}{\partial_+} \, t_2 \,
  \frac{1}{\partial_+} \, \ldots \, \frac{1}{\partial_+} \, t_2 \,
  \sim \, \delta (x^+).
\end{align}
Therefore $t_2 \, (1/\partial_+) \, \sim \, o(1)$ in $p_2^-$ power
counting.

Higher order corrections may come in through an insertion of one power
of $1/\partial_+^2$ between two $t_2$'s. One then gets
\begin{align}\label{Leik3}
  t_2 \, \frac{1}{\partial_+^2} \, t_2 \, \sim \, \delta (x^+) \, x^+
  \, \theta (x^+) \, = \, 0.
\end{align}
The equality in \eq{Leik3} is only true for delta-function shock waves
and demonstrates that \eq{Tmnd} is the exact solution for the problem
of the collision of two delta-function shock waves. For the smeared
shock waves of \eq{thetas} the zero in \eq{Leik3} would be replaces by
$a_2 \, \delta (x^+)$. As $a_2 \sim 1/p_2^-$ this indicates
suppression by a power of $1/p_2^-$ compared to the leading-order
terms in \eq{Leik2}. One can similarly show that insertions of higher
powers of $1/\partial_+$ would bring in further suppression. Thus our
approximation in Sect.  \ref{delta_sec} is justified by this explicit
power counting.

Let us now turn our attention to delta-prime shock waves of
\eq{tzero}. Notice that $t_1$ and $t_2$ in \eq{tzero} do not
explicitly depend on $p_1^+$ and $p_2^-$: as we show in Appendix
\ref{A} the dependence on these momenta (and, hence, on the
center-of-mass energy of the collision) comes in through singularities
like $\delta (x^\pm = 0)$. For $t_2 (x^+) \sim \delta' (x^+)$ one has
\begin{align}\label{Leik4}
  t_2 \, \frac{1}{\partial_+} \, t_2 \, \sim \, \delta' (x^+) \,
  \delta (x^+) \, \sim \left( \frac{\delta^2 (x^+)}{2} \right)' \,
  \sim \, p_2^- \, \delta' (x^+) .
\end{align}
Again we are not keeping track of factors not containing $p_2^-$.
Iterating the procedure we get
\begin{align}\label{Leik5}
  \left( t_2 \, \frac{1}{\partial_+} \right)^n \, t_2 \, \sim \, (p_2^-)^n \,
  \delta' (x^+) ,
\end{align}
that is, for delta-prime shock waves $t_2 \, (1/\partial_+) \, \sim
\, p_2^-$.

To understand higher order terms with more powers of $1/\partial_+$
consider
\begin{align}\label{Leik6}
  t_2 \, \frac{1}{\partial_+^2} \, t_2 \, \sim \, \delta' (x^+) \,
  \theta (x^+) \, \sim \, \left( \delta (x^+) \, \theta (x^+) \right)'
  - \delta^2 (x^+) \, \sim \, \left( \delta (x^+) \, \theta (x^+)
  \right)' - p_2^- \, \delta (x^+).
\end{align}
The term which was subleading for the delta-function shock waves (see
\eq{Leik3}) gives a leading-order factor of $p_2^-$ for delta-prime
shock waves, as we see from the last term in \eq{Leik6}.  Applying
higher powers of $t_2 \, (1/\partial_+)$ to the last term in
\eq{Leik6} does not make the term less important:
\begin{align}\label{Leik7}
  \left( t_2 \, \frac{1}{\partial_+} \right)^n \, p_2^- \, \delta
  (x^+) \, \sim \, (p_2^-)^{n+1} \, \delta (x^+).
\end{align}
(In fact the $\left( \delta (x^+) \, \theta (x^+) \right)'$ term in
\eq{Leik6} also brings in powers of $p_2^-$ after the operator $t_2 \,
(1/\partial_+)$ acts on it at least once.) We thus see that one
insertion of $t_2 (1/\partial_+^2)$ still gives leading terms in the
case of delta-prime shock waves. Fortunately higher order insertions
of $t_2 (1/\partial_+^2)$ start generating subleading terms and can be
discarded. We illustrate this by acting with $t_2 (1/\partial_+^2)$ on
the last term in \eq{Leik6}:
\begin{align}\label{Leik8}
  t_2 \, \frac{1}{\partial_+^2} \, p_2^- \, \delta (x^+) \, \sim \,
  p_2^- \, \delta' (x^+) \, x^+ \, \theta (x^+) \, \sim \, - p_2^- \,
  \delta (x^+) \, \theta (x^+).
\end{align}
No extra powers of $p_2^-$ is generated and hence such terms are
subleading. 

Insertions of a higher number of inverse derivatives are also
subleading. For instance
\begin{align}
  t_2 \, \frac{1}{\partial_+^3} \, t_2 \, \sim \, \delta' (x^+) \, \,
  x^+ \, \theta (x^+) \, \sim \, - \delta (x^+) \, \theta (x^+),
\end{align}
again producing no powers of $p_2^-$.

We conclude that for delta-prime shock waves the eikonal approximation
consists of the term in \eq{hbar4} along with all terms with a single
insertion of $t_2 (1/\partial_+^2)$ in all possible positions.


\subsubsection{Energy-Momentum Tensor of the Produced Medium}

To take into account all leading terms for the delta-prime shock waves
we write
\begin{align}
  {\bar h} \, = \, {\bar h}_{eik} + \delta {\bar h} 
\end{align}
with ${\bar h}_{eik}$ given by \eq{hbar4} and $\delta {\bar h}$
denoting the sum of all terms with all-orders of $t_2 (1/\partial_+)$
and exactly one insertion of $t_2 (1/\partial_+^2)$ as contained in
\eq{hbar3}.

Expanding \eq{hbar3} to the first order in $t_2 (1/\partial_+^2)$ we
write
\begin{align}\label{dh1}
  \delta {\bar h} \, = \, \left[ \left( 1 + \frac{1}{2} \, z^4 \, t_2
      \, \frac{\partial_-}{\partial_+} \right)^2 \right]^{-1} \,
  \left[ \frac{1}{2} \, z^4 \, t_2 \, \frac{1}{\partial_+^2} \, \left(
      {\hat D}_1 + {\hat D}_2 \right) + \frac{1}{8} \, z^4 \, t_2 \,
    \frac{\partial_-}{\partial_+} \, z^4 \, t_2 \,
    \frac{1}{\partial_+^2} \, {\hat D}_1 \right] \, {\bar h}_{eik}.
\end{align}
Expanding the first factor on the right hand side of \eq{dh1} into a
series and using the series representation for ${\bar h}_{eik}$ from
\eq{hbar5} yields
\begin{align}\label{dh2}
  \delta {\bar h} \, = \, \sum\limits_{m=0}^\infty \, (m+1) \, \left(
    - \frac{1}{2} \, z^4 \, t_2 \, \frac{\partial_-}{\partial_+}
  \right)^m \, \left[ \frac{1}{2} \, z^4 \, t_2 \,
    \frac{1}{\partial_+^2} \, \left( {\hat D}_1 + {\hat D}_2 \right) +
    \frac{1}{8} \, z^4 \, t_2 \, \frac{\partial_-}{\partial_+} \, z^4
    \, t_2 \, \frac{1}{\partial_+^2} \, {\hat D}_1 \right] \notag \\
  \times \, \sum\limits_{n=0}^\infty \, (n+1) \, \left( - \frac{1}{2}
    \, z^4 \, t_2 \, \frac{\partial_-}{\partial_+} \right)^n \, 16 \,
  z^7 \, t_1 \, t_2.
\end{align}
Using the definitions of ${\hat D}_1$ and ${\hat D}_2$ from Eqs.
(\ref{D1def}) and (\ref{D2def}) we obtain
\begin{align}
  {\hat D}_1 \, z^{4 \, n + 7} \, = \, - 8 \, (n+1) \, (2 \, n +1) \,
  z^{4 \, n + 5}
\end{align}
and
\begin{align}
  \left( {\hat D}_1 + {\hat D}_2 \right) \, z^{4 \, n + 7} \, = \, - 4
  \, (n+1) \, (2 \, n +3) \, z^{4 \, n + 5},
\end{align}
which allow us to rewrite \eq{dh2} as
\begin{align}\label{dh3}
  \delta {\bar h} \, = \, 16 \, \sum\limits_{n, m=0}^\infty \, (n+1)^2
  \, (m+1) \, \left( - \frac{1}{2} \right)^{n+m} \, \left( t_2 \,
    \frac{1}{\partial_+} \right)^m \, \left[ - 2 \, (2 \, n +3) \,
    z^{4 \, (n+m) + 9} \, t_2 \, \frac{1}{\partial_+^2} \right. \notag
  \\ \left. - (2 \, n +1) \, z^{4 \, (n+m) + 13} \, \partial_- \, t_2
    \, \frac{1}{\partial_+} \, t_2 \, \frac{1}{\partial_+^2} \right]
  \, \left[ \partial_-^{n+m} t_1 \right] \, \left( t_2 \,
    \frac{1}{\partial_+} \right)^n \, t_2.
\end{align}

The truncated amplitude contribution (\ref{dh3}) leads to the
contribution to the amplitude through \eq{hhbar}. Using the Green
function (\ref{G3}) we write
\begin{align}\label{dh4}
  \delta h \, = \, 2 \, z^2 \, \int\limits_{-\infty}^{x^+} d x'^+ \,
  \int\limits_{-\infty}^{x^-} d x'^- \, \int\limits_0^\infty d z' \,
  \int\limits_0^\infty d q \, q \, J_0\left( q \, \sqrt{2 \, (x^+ -
      x'^+) \, (x^- - x'^-)} \right) \, J_2 (q \, z) \, J_2 (q \, z')
  \notag \\ \times \, \sum\limits_{n, m=0}^\infty \, (n+1)^2 \, (m+1)
  \, \left( - \frac{1}{2} \right)^{n+m} \, \left( t_2 (x'^+) \,
    \frac{1}{\partial'_+} \right)^m \notag \\ \times \, \left[ -
    \frac{2 \, (2 \, n +3)}{n+m+2} \, (z')^{4 \, (n+m) + 7} \, t_2
    (x'^+) \, \frac{1}{\partial'^2_+} - \frac{2 \, n +1}{n+m+3} \,
    (z')^{4 \, (n+m) + 11} \, \partial'_- \, t_2 (x'^+) \,
    \frac{1}{\partial'_+} \, t_2 (x'^+)\, \frac{1}{\partial'^2_+}
  \right] \notag \\ \times \, \left[ \partial'^{n+m}_- t_1 (x'^-)
  \right] \, \left( t_2 (x'^+) \, \frac{1}{\partial'_+} \right)^n \,
  t_2 (x'^+).
\end{align}

As in Sect. \ref{delta_sec} we are interested in the contribution of
the metric element, now the term $\delta h$, to the transverse
pressure. As further evaluation of \eq{dh4} to all orders in $z$
appears to be rather involved, we expand it to the order $z^4$ using
\eq{Bess}, integrate over $z'$ and $q$, and eliminate the
$\partial'_-$ derivatives by successive integrations by parts. This
yields the following contribution to the transverse pressure
\begin{align}\label{dp1}
  \delta p \, = \, - \frac{N_c^2}{2 \, \pi^2} \, 8 & \,
  \int\limits_{-\infty}^{x^+} d x'^+ \, \int\limits_{-\infty}^{x^-} d
  x'^- \, \sum\limits_{n, m=0}^\infty \, (n+1)^2 \, (m+1) \, ( -
  2)^{n+m} \, (x^- - x'^-)^{n+m+2} \, t_1 (x'^-) \notag \\ & \times \,
  (x^+ - x'^+)^{2 \, (n+m) +2} \, \left[ 2 \, (2 \, n +3) \,
    \frac{\left( 2 \, (n+m) + 3 \right)!}{\left( n+m + 2 \right)!}
  \right. \notag \\ & \left.  + \, 4 \, (2 \, n + 1) \, \frac{\left( 2 \,
        (n+m) + 5 \right)!}{\left( n+m + 3 \right)!} \, (x^+ - x'^+)^2
    \, (x^- - x'^-) \, t_2 (x'^+) \, \frac{1}{\partial'_+} \right]
  \notag \\ & \times \, \left( t_2 (x'^+) \, \frac{1}{\partial'_+} \,
  \right)^m \, t_2 (x'^+)\, \frac{1}{\partial'^2_+} \, \left( t_2
    (x'^+) \, \frac{1}{\partial'_+} \right)^n \, t_2 (x'^+).
\end{align}

For further evaluation of \eq{dp1} we will explicitly substitute the
shock wave profiles from \eq{tzero}. As one can see from the
calculations in Appendix \ref{A} the regularization in \eq{tzero_reg}
is not needed for the leading-$p_2^-$ terms. 

First let us find the contribution to the transverse pressure coming
from the piece in \eq{peik}, which we will refer to as $p_{eik}$.  As
can be easily shown for $t_2$ from \eq{tzero} (see Appendix
\ref{dprimes})
\begin{align}\label{pile1}
  \left( t_2 (x^+) \, \frac{1}{\partial_+} \right)^n \, t_2 (x^+) \, =
  \, \frac{(\Lambda_2^2)^{n+1}}{(n+1)!} \, \left( \delta^{n+1} (x^+)
  \right)'. 
\end{align}
In \eq{pile1} and henceforth for simplicity we absorb factors of
$A_1^{1/3}$ and $A_2^{1/3}$ into $\Lambda_1^2$ and $\Lambda_2^2$.
Using Eqs.  (\ref{tzero}) and (\ref{pile1}) in \eq{peik} yields
\begin{align}\label{peik2}
  p_{eik} \, = \, \frac{N_c^2}{2 \, \pi^2} \, 8 \, \Lambda_1^2 \,
  \Lambda_2^2 \, \theta (x^+) \, \theta (x^-) \ \frac{1 - 44 \, p_2^-
    \, \Lambda_2^2 \, (x^+)^2 \, x^- + 64 \, \left( p_2^- \,
      \Lambda_2^2 \, (x^+)^2 \, x^- \right)^2 }{\left[ 1 + 8 \, p_2^-
      \, \Lambda_2^2 \, (x^+)^2 \, x^- \right]^{7/2}}.
\end{align}

To evaluate \eq{dp1} one needs another relation (see Appendix
\ref{dprimes} for its derivation), valid only for $t_2$ from
\eq{tzero} at the leading order in $p_2^-$
\begin{align}\label{pile2}
  \left( t_2 (x^+) \, \frac{1}{\partial_+} \, \right)^m \, t_2 (x^+)\,
  \frac{1}{\partial^2_+} \, \left( t_2 (x^+) \, \frac{1}{\partial_+}
  \right)^n \, t_2 (x^+) \, = \, \frac{(-1)^{m+1} \,
    (\Lambda_2^2)^{n+m+2} \, (p_2^-)^{n+m+1}}{(n+1)! \, (m+1)!} \,
  \delta (x^+).
\end{align}
Using Eqs. (\ref{tzero}) and (\ref{pile1}) in \eq{dp1} we get
\begin{align}\label{dp}
  \delta p \, = & \, \frac{N_c^2}{2 \, \pi^2} \, 8 \, \Lambda_1^2 \,
  \Lambda_2^2 \, \theta (x^+) \, \theta (x^-) \notag \\ & \times \,
  \left\{ 1 - 36 \, p_2^- \, \Lambda_2^2 \, (x^+)^2 \, x^- - \frac{1 -
      44 \, p_2^- \, \Lambda_2^2 \, (x^+)^2 \, x^- + 64 \, \left(
        p_2^- \, \Lambda_2^2 \, (x^+)^2 \, x^- \right)^2 }{\left[ 1 +
        8 \, p_2^- \, \Lambda_2^2 \, (x^+)^2 \, x^- \right]^{7/2}}
\right\}.
\end{align}

The net transverse pressure is obtained by adding Eqs. (\ref{peik2})
and (\ref{dp})
\begin{align}\label{pdp}
  p \, = \, p_{eik} + \delta p \, = \, \frac{N_c^2}{2 \, \pi^2} \, 8
  \, \Lambda_1^2 \, \Lambda_2^2 \, \theta (x^+) \, \theta (x^-) \ 
  \left[ 1 - 36 \, p_2^- \, \Lambda_2^2 \, (x^+)^2 \, x^- \right].
\end{align}
This is the first main result of this Section. Importantly all higher
order terms cancel leaving us with the simple expression (\ref{pdp})!
Note that \eq{pdp} is just a sum of the LO and NLO corrections
(resumming leading powers of $p_2^-$), and thus it agrees with
\eq{ppNLO}. Namely it turns out that the NLO transverse pressure from
\eq{ppNLO} taken at the leading-$p_2^-$ accuracy gives us the full
eikonal result for the proton-nucleus scattering problem with
delta-prime shock waves.

\eq{pdp} indeed has a limited region of applicability. As it was
derived for the proton-nucleus approximation, similar to \eq{bounds}
we must have
\begin{align}
  p_1^+ \, \Lambda_1^2 \, (x^-)^2 \, x^+ \, \ll \, 1, \ \ \ p_2^- \,
  \Lambda_2^2 \, (x^+)^2 \, x^- \, \sim 1,
\end{align} 
to be able to neglect eikonal graviton exchanges with the proton shock
wave. We also want to neglect the non-eikonal terms shown in
\eq{ppNLO}, which requires (see \eq{bounds2})
\begin{align}
  \Lambda_1^2 \, \tau^2 \, \ll \, 1, \ \ \ \Lambda_2^2 \, \tau^2 \,
  \ll \, 1.
\end{align}

It appears that the region of applicability of \eq{pdp} is indeed
somewhat limited and is confined to the region of large $x^+$ and
small $x^-$, i.e., the region of space-time in the forward light cone
bordering the proton shock wave. Still it is perhaps surprising to see
that the pressure in \eq{pdp} can easily become negative at large
enough $x^+$. As the pressure is negative we conclude that the system
has not yet reached the ideal hydrodynamics state, similar to what
happened to the delta-function shock waves in Sect. \ref{delta_sec}.

The presence of negative pressure does not pose any problems by
itself: negative pressure is known to arise in the early stages of
heavy ion collisions when they are described in the Color Glass
Condensate framework \cite{Krasnitz:2003jw,Krasnitz:2002mn}. Even in
the strongly-coupled theory considered here, the LO part of the
transverse pressure (\ref{ppNLO}), when used in \eq{prod_Tmn2}, leads
to negative pressure in the longitudinal direction
\cite{Albacete:2008vs}. In comparison, appearance of a negative energy
density would be indeed worrisome and would indicate an unphysical
situation.  However, we can not calculate the energy density here, as
the matter distribution is indeed rapidity-dependent and is not
described by the ideal hydrodynamics: it is impossible to find the
local rest frame of such medium to meaningfully talk about the energy
density.

Therefore negative pressure in \eq{pdp} may be physical. One could
interpret it as follows: when we chose the delta-prime shock waves of
\eq{tzero} we ``forced'' the shock waves not to stop and to continue
along the light cone trajectories. At the same time the produced
strongly-interacting medium is still trying to pull them back
together. As the shock waves are ``artificially'' pinned down to their
light cones they do not stop, thus creating a negative pressure in the
medium which tries to slow them down.

Using Eqs. (\ref{pdp}), (\ref{pdef}) and (\ref{prod_Tmn2}) we
construct all non-zero components of the energy-momentum tensor of the
produced medium:
\begin{subequations}\label{Tmnp}
\begin{align}
  \langle T^{++}\rangle \, & = \, \frac{N_c^2}{2 \, \pi^2} \, \left[ -
    8 \, \Lambda_1^2 \, \Lambda_2^2 \, \delta (x^-) \, x^+ \, \theta
    (x^+) + 96 \, p_2^- \, \Lambda_1^2 \, \Lambda_2^4 \, (x^+)^3 \,
    \theta (x^+) \, \theta (x^-)  \right], \\
  \langle T^{--}\rangle \, & = \, \frac{N_c^2}{2 \, \pi^2} \, \left[ -
    8 \, \Lambda_1^2 \, \Lambda_2^2 \, \delta (x^+) \, x^- \, \theta
    (x^-) + 288 \, p_2^- \, \Lambda_1^2 \, \Lambda_2^4 \ x^+ \,
    \theta (x^+) \, (x^-)^2 \, \theta (x^-) \right], \\
  \langle T^{+-}\rangle \, & = \, \frac{N_c^2}{2 \, \pi^2} \ 8 \,
  \Lambda_1^2 \, \Lambda_2^2 \, \theta (x^+) \, \theta (x^-) \ \left[
    1 - 36 \, p_2^- \, \Lambda_2^2 \, (x^+)^2 \, x^- \right], \\
  \langle T^{\, i j}\rangle \, & = \, \delta^{ij} \, \frac{N_c^2}{2
    \,\pi^2} \ 8 \, \Lambda_1^2 \, \Lambda_2^2 \, \theta (x^+) \,
  \theta (x^-) \ \left[ 1 - 36 \, p_2^- \, \Lambda_2^2 \, (x^+)^2 \,
    x^- \right].
\end{align}
\end{subequations}
This is the second main result of this Section.


\subsection{Validity Range of the Perturbative Expansion}

Before proceeding to the conclusions, let us check the validity range
of the perturbative approach for solving Einstein equations followed
throughout the paper and outlined in \eq{texp}.  For the sake of
simplicity, we will explicitly analyze the relative contribution to
the $\perp\perp$ metric coefficient obtained at LO ($H^{(0)}$), and
NLO ($H^{(1)}$), for the case of (physical) delta function shock waves
of \eq{deltas}. Starting from \eq{LO} for the LO results and Eqs.
(\ref{H}-\ref{H1coefs}) for the NLO one and dropping some trivial
factors of order one, the ratio $R_{NLO/LO}$ between the NLO and LO
contribution to the metric coefficient $H(x^+,x^-,z)$ is given by
\begin{align}\label{R}
  R_{NLO/LO}\equiv\frac{H^{(1)}(x^+,x^-,z)}{H^{(0)}(x^+,x^-,z)} \sim
  \mu_2 \, \tau^3 \,
  e^{\eta}\,\frac{1+(z/\tau)^2+(z/\tau)^4+(z/\tau)^6}{1+(z/\tau)^2}.
\end{align}
For definitiveness we have chosen to concentrate on the higher order
corrections due to graviton exchanges with the shock wave described by
the energy scale $\mu_2$. To obtain an estimate for the NLO graviton
exchanges in the other shock wave one simply has to replace $\mu_2
\rightarrow \mu_1$ and $\eta \rightarrow - \eta$ in \eq{R}. In
arriving at \eq{R} we have made use of the relation $x^+ =
\tau\,e^{\eta}/\sqrt{2}$, with $\eta = (1/2) \ln (x^+/x^-)$ the
space-time rapidity. Note that all the coefficients in the numerator
and denominator in the ratio in \eq{R} are put to be equal to $1$ for
simplicity of the parametric estimate we are performing. 

\FIGURE[htb]{\includegraphics[width=10cm,angle=270]{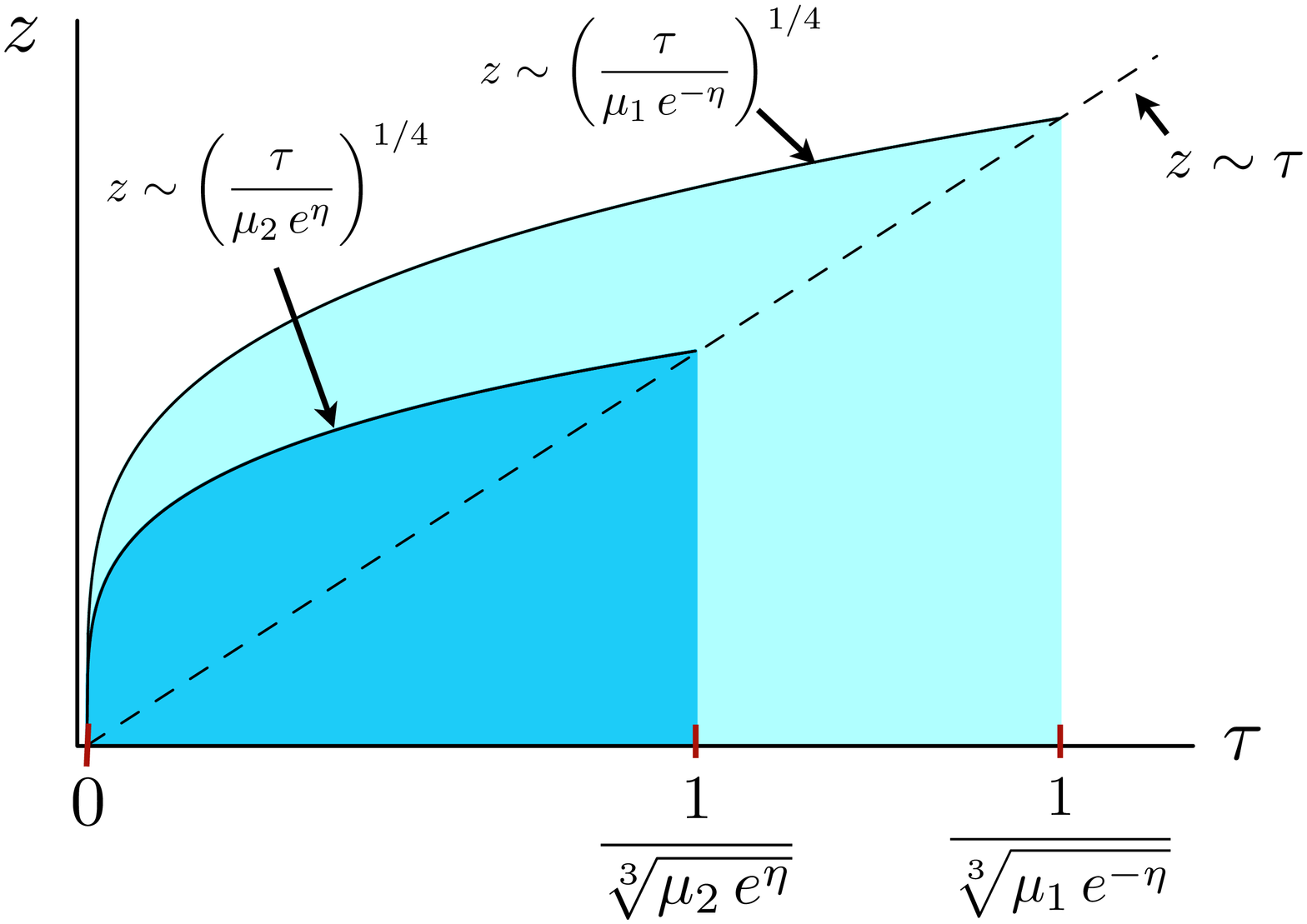}
  \caption{Schematic plot of the validity range of our solution in the
    $z$-$\tau$ plane. The $z$-axis was moved to a slight left of $\tau
    =0$ line for illustrative purposes only: indeed $\tau \ge 0$. The
    darker shaded area indicated the validity region of the
    perturbative expansion of Sect. \ref{pert}. The lighter shaded
    area (together with the darker one) depicts the broader validity
    region of the pA approximation developed in this Section.}
  \label{range}
}

Similar to \eq{R}, one can build the ratio of the NNLO contribution,
$H^{(2)}$ given by Eqs. (\ref{psis}--\ref{H2}), to the NLO one
$H^{(1)}$, again for the delta-function shock waves:
\begin{align}\label{R2}
  R_{NNLO/NLO} \equiv \frac{H^{(2)}(x^+,x^-,z)}{H^{(1)}(x^+,x^-,z)}
  \sim \mu_2 \, \tau^3 \, e^{\eta} \,
  \frac{1+(z/\tau)^2+(z/\tau)^4+(z/\tau)^6}{1+(z/\tau)^2} \sim
  R_{NLO/LO}.
\end{align}

One can easily show that Eqs. (\ref{R}) and (\ref{R2}) are also valid
for delta-prime shock waves of \eq{tzero}. Thus, the condition
$R_{NLO/LO} \lesssim 1$ sets the validity range of the whole
perturbative expansion.  Two different situations can be considered:
($i$) $z \ll \tau$: in this case the perturbative expansion is
justified as long as $\tau \lesssim [1/(\mu_2 \, e^{\eta})]^{1/3}$.
($ii$) $z\gg \tau$: in this case $R_{NLO/LO} \lesssim 1$ if $z
\lesssim [\tau/(\mu_2\,e^{\eta})]^{1/4}$. The validity region of our
perturbative expansion given by the union of regions ($i$) and ($ii$)
is depicted in \fig{range} (see the darker shaded region there). Thus
our approximation is valid in comparable intervals in $\tau$ and $z$
near the boundary of the AdS space.

On the other hand, in the calculation performed in this Section for
asymmetric pA collisions, we resummed {\sl all-order} graviton
exchanges with the nucleus shock wave. Therefore the applicability of
this approximation does not require $R_{NLO/LO} \lesssim 1$ for higher
order corrections bringing in powers of $\mu_2$. We do neglect all
higher-order graviton exchanges with the proton shock wave that bring
in higher powers of $\mu_1$. Thus our pA approximation is valid only
if $R_{NLO/LO} \lesssim 1$ for NLO corrections with $\mu_2 \rightarrow
\mu_1$ and $\eta \rightarrow - \eta$ in \eq{R}. The two regions ($i$)
and ($ii$) become: ($i$) $z \ll \tau$, $\tau \lesssim [1/(\mu_1 \,
e^{-\eta})]^{1/3}$, and ($ii$) $z\gg \tau$, $z \lesssim [\tau/(\mu_1
\, e^{-\eta})]^{1/4}$. These new regions ($i$) and ($ii$) for pA
collisions are shown in \fig{range} by the lightly shaded area. Indeed
the validity of the pA approximation is much broader than that of the
validity of the perturbation series of Sect. \ref{pert}.


Importantly, the stopping time estimated in Sect. \ref{prot-stop},
$\tau_{stop} \sim [1/(\mu_2e^{\eta})]^{1/3}$, lies within the validity
range of our approximation, as one can easily see from \fig{range}.
Our approach is valid for a comparably broad range of $z$'s at the
stopping time, though for the physics of the gauge theory in four
dimensions only the knowledge of the metric in the small-$z$ region is
needed.


\section{Conclusions}
\label{conc}

Let us summarize our main results. In Sect. \ref{pert} we constructed
the NLO and NNLO terms in the perturbative expansion in graviton
exchanges for the collision of two shock waves. Our expansion
generalizes similar expansion constructed previously in
\cite{Grumiller:2008va} from the delta-function-only case considered
in \cite{Grumiller:2008va} to the case of shock waves of arbitrary
profile. In particular we see that even the delta-prime shock waves,
which at LO gave rapidity-independent distribution of matter
\cite{Albacete:2008vs}, lead to rapidity-dependent energy-momentum
tensor at NLO (see \eq{ppNLO}).

It is worthwhile noting that the perturbative graviton expansion of
Sect. \ref{pert} can be built consistently without introducing a
dilaton field. Hence at all orders of the solution the dilaton field
is zero. As the dilaton field is dual to the $\langle \mbox{tr}
F_{\mu\nu}^2 \rangle$ operator in the gauge theory at the boundary we
conclude that for the collisions of shock waves considered here
\begin{align}
  \langle \mbox{tr} F_{\mu\nu}^2 \rangle \, = \, 0
\end{align}
at all times. Thus electric and magnetic modes are always equilibrated
in this strongly-coupled medium. This result should be contrasted with
that of \cite{Heller:2007qt}, where dilaton field was needed to
construct singularity-free pre-asymptotics to the Bjorken
hydrodynamics metric of \cite{Janik:2005zt}. As our calculations show,
the absence of a dilaton field in the initial shock waves leads to no
dilaton field throughout the collision. As it is difficult to
construct shock waves with non-zero dilaton field it is not clear how
to construct shock wave collisions with non-zero dilaton field in the
forward light cone. Therefore the no-dilaton aspect of our result may
give one reasons to worry whether dual-Bjorken geometry of
\cite{Janik:2005zt} is obtainable at all in collisions of AdS shock
waves.

In Sect. \ref{pA} we have devised an eikonal resummation procedure,
which resumed all graviton rescattering in one nucleus while keeping
only one graviton exchange with another nucleus. The results for
delta-function shock waves are given in \eq{Tmnd}. As is clear from
\eq{Tmnd} the matter distribution obtained in the proton-nucleus
approximation can not be described by ideal hydrodynamics, and should
be viewed as some intermediate stage of the matter evolution towards
isotropization. We also showed explicitly in Sect. \ref{pA} that
strong interactions with the nucleus stop the proton completely, as
can be seen from \eq{stop}.

The eikonal expansion for an asymmetric collision of two delta-prime
shock waves terminates at the level of two graviton-exchange with the
nucleus. The results of the resummation are shown in \eq{Tmnp}. It is
important to note that the energy-momentum tensors for delta-function
shock waves (\ref{Tmnd}) and for delta-primes (\ref{Tmnp}) are
strongly rapidity-dependent. It is unlikely that a matter distribution
which is strongly rapidity-dependent at early times would become
rapidity-independent at late times: such behavior would be acausal, as
different rapidity regions become causally disconnected from each
other as the collision evolves.  It is therefore probable that
collisions of shock waves in AdS will lead to a rapidity-dependent
final state at late times: if, due to strong interactions, the matter
in this late-time state would be described by the ideal hydrodynamics,
this hydrodynamic description can not be that of rapidity-independent
Bjorken hydrodynamics \cite{Bjorken:1982qr}.  While a full (possibly
numerical) study of the nucleus-nucleus collision in AdS would provide
definitive answer to this question, it is possible that Bjorken
geometry gives a good approximation to the dynamics of the matter
produced in a collision of two identical nuclei only in a narrow
interval around mid-rapidity. Our results here can serve as a
benchmark for further (possibly numerical) studies of the collision of
two shock waves beyond the asymmetric approximation done here.

It may be that to obtain Bjorken hydrodynamics in a broader rapidity
range one has to abandon the idea of colliding shock waves in AdS, and
try to simulate the initial stage of the medium by matching the AdS
metric onto the results for the energy-momentum tensor known from
weak-coupling CGC methods
\cite{Krasnitz:2003jw,Krasnitz:2002mn,Lappi:2003bi,Fukushima:2007ja,Kovchegov:2007pq,Kovchegov:2005ss}.
Such approach was advocated in
\cite{Chesler:2008hg,Lin:2009pn,Kovchegov:2007pq} and may prove to be
quite fruitful. One possible shortcoming of the matching method is in
the fact that it leaves too much freedom in the choice of the
early-time AdS metric, leading to a possible loss of uniqueness in the
description of the subsequent time-evolution of the system. Further
research is needed to understand which AdS approach is better suited
to describe heavy ion collisions.


\acknowledgments

We would like to thank Samir Mathur and Robert Myers for informative
discussions.

The work of Yu.K. and A.T. is sponsored in part by the U.S. Department
of Energy under Grant No. DE-FG02-05ER41377.


\appendix

\renewcommand{\theequation}{A\arabic{equation}}
  \setcounter{equation}{0}
\section{Transverse Pressure for Delta-Primes at NLO}
\label{A}

We want to find the NLO contribution to the transverse pressure due to
delta-prime sources. Using \eq{pexp} we write
\begin{align}\label{p1}
  p_{NLO} \, = \, \frac{N_c^2}{2 \, \pi^2} \, H_0^{(1)} (x^+, x^-) \,
  = \, \frac{N_c^2}{2 \, \pi^2} \left[ - \frac{6}{(\partial_{+} \,
      \partial_{-})^{2}} \, \psi_{7} - \frac{96}{(\partial_{+} \,
      \partial_{-})^{3}} \, \psi_{9} - \frac{2880}{(\partial_{+} \,
      \partial_{-})^{4}} \, \psi_{11} \right]
\end{align}
where we used $H_0^{(1)}$ given by \eq{H10}. Our goal is to evaluate
the right hand side of \eq{p1} for $t_1$ and $t_2$ given by
\eq{tzero_reg}. For simplicity we will evaluate the terms with one
power of $t_1$ only: the remaining terms with only one power of $t_2$
can be obtained by the substitution $t_1 \leftrightarrow t_2$.

Using Eqs. (\ref{ps}) along with Eqs. (\ref{LOstuff}) we write
\begin{align}\label{Aterm}
  - \frac{6}{(\partial_{+} \, \partial_{-})^{2}} \, \psi_{7} \, = \, -
  64 \, \left[ \frac{1}{\partial_-^3} t_1 (x^-) \right] \, \left[
    \frac{1}{\partial_+^2} t_2 (x^+) \, \frac{1}{\partial_+^3} t_2
    (x^+) \right] + (t_1 \leftrightarrow t_2).
\end{align}
Using \eq{tzero_reg} we get
\begin{align}\label{mint}
  \frac{1}{\partial_-^3} t_1 (x^-) \, = \, \Lambda_1^2 \,
  \sum\limits_{i=1}^{A_1^{1/3}} \, (x^- - x_i^-) \, \theta (x^- -
  x_i^-) \, \approx \, \Lambda_1^2 \, A_1^{1/3} \, x^- \, \theta (x^-)
\end{align}
for $x^- \gg a_1$. Similarly we write
\begin{align}\label{pint}
  \frac{1}{\partial_+^2} t_2 (x^+) \, & \frac{1}{\partial_+^3} t_2
  (x^+) \, = \, \Lambda_2^4 \, \sum\limits_{i,j=1}^{A_2^{1/3}} \,
  \int\limits_{-\infty}^{x^+} d x'^+ \, \int\limits_{-\infty}^{x'^+} d
  x''^+ \, \delta' (x''^+ - x^+_i) \, (x''^+ - x^+_j) \, \theta (x''^+
  - x^+_j) \notag \\
  & = \Lambda_2^4 \, \sum\limits_{i,j=1}^{A_2^{1/3}} \,
  \int\limits_{-\infty}^{x^+} d x'^+ \, \left[ \delta (x'^+ - x^+_i)
    \, (x_i^+ - x^+_j) \, \theta (x_i^+ - x^+_j) - \theta (x'^+ -
    x^+_i) \, \theta (x_i^+ - x^+_j) \right] \notag \\ & \approx
  \frac{(\Lambda_2^2 \, A_2^{1/3})^2}{2} \, x^+ \, \theta (x^+),
\end{align}
where in the last step we have used $x^+ \gg a_2$. Combining Eqs.
(\ref{mint}) and (\ref{pint}) we obtain
\begin{align}\label{psi7}
  - \frac{6}{(\partial_{+} \, \partial_{-})^{2}} \, \psi_{7} \,
  \approx \, - 32 \, \Lambda_1^2 \, A_1^{1/3} \, (\Lambda_2^2 \,
  A_2^{1/3})^2 \, x^+ \, x^- \, \theta (x^+) \, \theta (x^-) + (1
  \leftrightarrow 2).
\end{align}

Eqs. (\ref{ps}), (\ref{LOstuff}) give
\begin{align}\label{Bterm}
  - \frac{96}{(\partial_{+} \, \partial_{-})^{3}} \, \psi_{9} \, = \,
  - 576 \, \left[ \frac{1}{\partial_-^3} t_1 (x^-) \right] \, \left[
    \frac{1}{\partial_+^3} t_2 (x^+) \, \frac{1}{\partial_+^2} t_2
    (x^+) \right] + (t_1 \leftrightarrow t_2).
\end{align}
The only difference of \eq{Bterm} with \eq{Aterm} is in
$t_2$-dependent part, which is evaluated to give
\begin{align}\label{step1}
  & \frac{1}{\partial_+^3} t_2 (x^+) \, \frac{1}{\partial_+^2} t_2
  (x^+) \, = \, \Lambda_2^4 \, \sum\limits_{i,j=1}^{A_2^{1/3}} \,
  \int\limits_{-\infty}^{x^+} d x'^+ \, \int\limits_{-\infty}^{x'^+} d
  x''^+ \, \int\limits_{-\infty}^{x''^+} d x'''^+ \, \delta' (x'''^+ -
  x^+_i) \, \theta (x'''^+ - x^+_j) \notag \\ & = \Lambda_2^4
  \sum\limits_{i,j=1}^{A_2^{1/3}} \int\limits_{-\infty}^{x^+} d x'^+
  \int\limits_{-\infty}^{x'^+} d x''^+ \int\limits_{-\infty}^{x''^+} d
  x'''^+ \, \left[ \partial'''_+ \left( \delta (x'''^+ - x^+_i) \,
      \theta (x'''^+ - x^+_j) \right) - \delta (x'''^+ - x^+_i) \,
    \delta (x'''^+ - x^+_j) \right].
\end{align}
The last term in the last line of \eq{step1} is only non-zero when
$x_i^+ = x_j^+$, which is only true if $i=j$ as all the $x_i^+$'s are
different. However, if $i=j$ that term becomes a delta-function
squared. Regulating the infinity by the largest momentum scale in the
problem we replace $\delta (x^+ = 0) \rightarrow p_2^-$ and get
\begin{align}\label{step2}
  \frac{1}{\partial_+^3} t_2 (x^+) \, \frac{1}{\partial_+^2} t_2 (x^+)
  \, & = \, \Lambda_2^4 \, \sum\limits_{i,j=1}^{A_2^{1/3}}
  \int\limits_{-\infty}^{x^+} d x'^+ \int\limits_{-\infty}^{x'^+} d
  x''^+ \, \left[ \delta (x''^+ - x^+_i) \, \theta (x_i^+ - x^+_j) -
    \delta_{i j} \, p_2^- \, \theta (x''^+ - x_i^+) \right] \notag \\
  & \approx \, \frac{(\Lambda_2^2 \, A_2^{1/3})^2}{2} \, x^+ \, \theta
  (x^+) \, \left[ 1 - \frac{p_2^- \, x^+}{A_2^{1/3}} \right]
\end{align}
for $x^+ \gg a_2$. Combining Eqs. (\ref{mint}) and (\ref{step2}) in
\eq{Bterm} we get
\begin{align}\label{psi9}
  - \frac{96}{(\partial_{+} \, \partial_{-})^{3}} \, \psi_{9} \, = \,
  - 288 \, \Lambda_1^2 \, A_1^{1/3} \, (\Lambda_2^2 \, A_2^{1/3})^2 \,
  x^+ \, x^- \, \theta (x^+) \, \theta (x^-) \, \left[ 1 - \frac{p_2^-
      \, x^+}{A_2^{1/3}} \right] + (1 \leftrightarrow 2, \, +
  \leftrightarrow - ).
\end{align}
(Indeed the delta function $\delta (x^-=0)$ should be regulated by
$p_1^+$.)

Similar to the above one gets
\begin{align}\label{psi11}
  - \frac{2880}{(\partial_{+} \, \partial_{-})^{4}} \, \psi_{11} \, =
  \, - 2304 \, \left[ \frac{1}{\partial_-^3} t_1 (x^-) \right] \,
  \left[ \frac{1}{\partial_+^4} t_2 (x^+) \, \frac{1}{\partial_+} t_2
    (x^+) \right] + (t_1 \leftrightarrow t_2) \notag \\
  \, = \, - 576 \, \Lambda_1^2 \, A_1^{1/3} \, (\Lambda_2^2 \,
  A_2^{1/3})^2 \, x^+ \, x^- \, \theta (x^+) \, \theta (x^-) \,
  \frac{p_2^- \, x^+}{A_2^{1/3}} + (1 \leftrightarrow 2, \, +
  \leftrightarrow - ).
\end{align}

Using Eqs. (\ref{psi7}), (\ref{psi9}), and (\ref{psi11}) in \eq{p1}
yields the NLO contribution to the pressure given in \eq{ppNLO}.


\renewcommand{\theequation}{B\arabic{equation}}
  \setcounter{equation}{0}
\section{Iterations of Delta-Primes}
\label{dprimes}

We start by proving \eq{pile1} for 
\begin{align}\label{t2zero}
  t_2 (x^+) \, = \, \Lambda_2^2 \, \delta' (x^+). 
\end{align}
Evaluating one iteration of $t_2 (1/\partial_+)$ operator acting on
$t_2$ we get
\begin{align}
  t_2 (x^+) \, \frac{1}{\partial_+} \, t_2 (x^+) \, = \,
  (\Lambda_2^2)^2 \, \delta' (x^+) \, \delta (x^+) \, =
  (\Lambda_2^2)^2 \, \left( \frac{\delta^2 (x^+)}{2} \right)'. 
\end{align}
Similarly
\begin{align}
  \left( t_2 (x^+) \, \frac{1}{\partial_+} \right)^2 \, t_2 (x^+) \, =
  \, (\Lambda_2^2)^3 \, \delta' (x^+) \, \frac{\delta^2 (x^+)}{2} \, =
  \, (\Lambda_2^2)^3 \, \left( \frac{\delta^3 (x^+)}{3!} \right)'.
\end{align}
It is now straightforward to see what happens at each step of
application of the $t_2 (1/\partial_+)$ operator to write
\begin{align}\label{pile10}
  \left( t_2 (x^+) \, \frac{1}{\partial_+} \right)^n \, t_2 (x^+) \, =
  \, \frac{(\Lambda_2^2)^{n+1}}{(n+1)!} \, \left( \delta^{n+1} (x^+)
  \right)', 
\end{align}
which is exactly \eq{pile1}, as desired.

Now let us prove \eq{pile2} for $t_2$ from \eq{t2zero}. First of all
use \eq{pile10} that we just proved to write
\begin{align}\label{pile20}
  \left( t_2 (x^+) \, \frac{1}{\partial_+} \, \right)^m \, t_2 (x^+)\,
  \frac{1}{\partial^2_+} \, \left( t_2 (x^+) \, \frac{1}{\partial_+}
  \right)^n \, t_2 (x^+) \, = \, \frac{(\Lambda_2^2)^{n+1}}{(n+1)!} \,
  \left( t_2 (x^+) \, \frac{1}{\partial_+} \, \right)^{m+1} \,
  \delta^{n+1} (x^+).
\end{align}
To evaluate 
\begin{align}
  \left( t_2 (x^+) \, \frac{1}{\partial_+} \, \right)^{m+1} \,
  \delta^{n+1} (x^+)
\end{align}
we write
\begin{align}\label{longpile}
  \left( \delta' (x^+) \, \frac{1}{\partial_+} \right)^{m+1} \, & = \,
  \delta' (x^+) \, \frac{1}{\partial_+} \, \delta' (x^+) \,
  \frac{1}{\partial_+} \, \ldots \, \delta' (x^+) \,
  \frac{1}{\partial_+} \, = - \left( \frac{\delta^2 (x^+)}{2} \right)'
  \, \frac{1}{\partial_+} \, \left( \delta' (x^+) \,
    \frac{1}{\partial_+} \right)^{m-1} \notag \\
  & = \, \left( \frac{\delta^3 (x^+)}{3!} \right)' \,
  \frac{1}{\partial_+} \, \left( \delta' (x^+) \, \frac{1}{\partial_+}
  \right)^{m-2} \, = \, \ldots \, = \, (-1)^m \, \left(
    \frac{\delta^{m+1} (x^+)}{(m+1)!} \right)' \,
  \frac{1}{\partial_+}.
\end{align}
In each step in \eq{longpile} we neglected a total derivative: it can
be shown that those total derivatives do not generate leading powers
of $p_2^-$. \eq{longpile} gives
\begin{align}\label{longpile2}
  \left( t_2 (x^+) \, \frac{1}{\partial_+} \, \right)^{m+1} \,
  \delta^{n+1} (x^+) \, = \, (\Lambda_2^2)^{m+1} \, (-1)^m \, \left(
    \frac{\delta^{m+1} (x^+)}{(m+1)!} \right)' \, \frac{1}{\partial_+}
  \, \delta^{n+1} (x^+) \notag \\ = \, \frac{(-1)^{m+1} \,
    (\Lambda_2^2)^{m+1}}{(m+1)!} \, \delta^{n+m+2} (x^+) \, = \, \frac{(-1)^{m+1} \,
    (\Lambda_2^2)^{m+1} \, (p_2^-)^{n+m+1}}{(m+1)!} \, \delta (x^+),
\end{align}
where we again neglected the total derivative as it is subleading. In
\eq{longpile2} we have also regularized the extra powers of $\delta
(x^+ =0)$ by replacing them with $p_2^-$.

Combining Eqs. (\ref{longpile2}) and (\ref{pile20}) yields
\begin{align}\label{pile21}
  \left( t_2 (x^+) \, \frac{1}{\partial_+} \, \right)^m \, t_2 (x^+)\,
  \frac{1}{\partial^2_+} \, \left( t_2 (x^+) \, \frac{1}{\partial_+}
  \right)^n \, t_2 (x^+) \, = \, \frac{(-1)^{m+1} \,
    (\Lambda_2^2)^{n+m+2} \, (p_2^-)^{n+m+1}}{(n+1)! \ (m+1)!} \,
  \delta (x^+),
\end{align}
which is exactly \eq{pile2}.



\begin{thebibliography}{10}

\bibitem{Albacete:2008vs}
J.~L. Albacete, Y.~V. Kovchegov, and A.~Taliotis, {\it {Modeling Heavy Ion
  Collisions in AdS/CFT}},  {\em JHEP} {\bf 07} (2008) 100,
  [\href{http://xxx.lanl.gov/abs/0805.2927}{{\tt arXiv:0805.2927}}].

\bibitem{Maldacena:1997re}
J.~M. Maldacena, {\it The large n limit of superconformal field theories and
  supergravity},  {\em Adv. Theor. Math. Phys.} {\bf 2} (1998) 231--252,
  [\href{http://xxx.lanl.gov/abs/hep-th/9711200}{{\tt hep-th/9711200}}].

\bibitem{Gubser:1998bc}
S.~S. Gubser, I.~R. Klebanov, and A.~M. Polyakov, {\it Gauge theory correlators
  from non-critical string theory},  {\em Phys. Lett.} {\bf B428} (1998)
  105--114, [\href{http://xxx.lanl.gov/abs/hep-th/9802109}{{\tt
  hep-th/9802109}}].

\bibitem{Witten:1998qj}
E.~Witten, {\it Anti-de sitter space and holography},  {\em Adv. Theor. Math.
  Phys.} {\bf 2} (1998) 253--291,
  [\href{http://xxx.lanl.gov/abs/hep-th/9802150}{{\tt hep-th/9802150}}].

\bibitem{Aharony:1999ti}
O.~Aharony, S.~S. Gubser, J.~M. Maldacena, H.~Ooguri, and Y.~Oz, {\it Large n
  field theories, string theory and gravity},  {\em Phys. Rept.} {\bf 323}
  (2000) 183--386, [\href{http://xxx.lanl.gov/abs/hep-th/9905111}{{\tt
  hep-th/9905111}}].

\bibitem{D'Eath:1992hb}
P.~D. D'Eath and P.~N. Payne, {\it {Gravitational radiation in high speed black
  hole collisions. 1. Perturbation treatment of the axisymmetric speed of light
  collision}},  {\em Phys. Rev.} {\bf D46} (1992) 658--674.

\bibitem{D'Eath:1992hd}
P.~D. D'Eath and P.~N. Payne, {\it {Gravitational radiation in high speed black
  hole collisions. 2. Reduction to two independent variables and calculation of
  the second order news function}},  {\em Phys. Rev.} {\bf D46} (1992)
  675--693.

\bibitem{D'Eath:1992qu}
P.~D. D'Eath and P.~N. Payne, {\it {Gravitational radiation in high speed black
  hole collisions. 3. Results and conclusions}},  {\em Phys. Rev.} {\bf D46}
  (1992) 694--701.

\bibitem{Sperhake:2008ga}
U.~Sperhake, V.~Cardoso, F.~Pretorius, E.~Berti, and J.~A. Gonzalez, {\it {The
  high-energy collision of two black holes}},  {\em Phys. Rev. Lett.} {\bf 101}
  (2008) 161101, [\href{http://xxx.lanl.gov/abs/0806.1738}{{\tt
  arXiv:0806.1738}}].

\bibitem{Nastase:2005rp}
H.~Nastase, {\it {The RHIC fireball as a dual black hole}},
  \href{http://xxx.lanl.gov/abs/hep-th/0501068}{{\tt hep-th/0501068}}.

\bibitem{Kajantie:2008rx}
K.~Kajantie, J.~Louko, and T.~Tahkokallio, {\it {Gravity dual of conformal
  matter collisions in 1+1 dimensions}},  {\em Phys. Rev.} {\bf D77} (2008)
  066001, [\href{http://xxx.lanl.gov/abs/0801.0198}{{\tt arXiv:0801.0198}}].

\bibitem{Grumiller:2008va}
D.~Grumiller and P.~Romatschke, {\it {On the collision of two shock waves in
  AdS5}},  {\em JHEP} {\bf 08} (2008) 027,
  [\href{http://xxx.lanl.gov/abs/0803.3226}{{\tt arXiv:0803.3226}}].

\bibitem{Gubser:2008pc}
S.~S. Gubser, S.~S. Pufu, and A.~Yarom, {\it {Entropy production in collisions
  of gravitational shock waves and of heavy ions}},  {\em Phys. Rev.} {\bf D78}
  (2008) 066014, [\href{http://xxx.lanl.gov/abs/0805.1551}{{\tt
  arXiv:0805.1551}}].

\bibitem{Lin:2009pn}
S.~Lin and E.~Shuryak, {\it {Grazing Collisions of Gravitational Shock Waves
  and Entropy Production in Heavy Ion Collision}},
  \href{http://xxx.lanl.gov/abs/0902.1508}{{\tt arXiv:0902.1508}}.

\bibitem{Kolb:2000sd}
P.~F. Kolb, J.~Sollfrank, and U.~W. Heinz, {\it Anisotropic transverse flow and
  the quark-hadron phase transition},  {\em Phys. Rev.} {\bf C62} (2000)
  054909, [\href{http://xxx.lanl.gov/abs/hep-ph/0006129}{{\tt
  hep-ph/0006129}}].

\bibitem{Kolb:2000fh}
P.~F. Kolb, P.~Huovinen, U.~W. Heinz, and H.~Heiselberg, {\it {Elliptic flow at
  SPS and RHIC: From kinetic transport to hydrodynamics}},  {\em Phys. Lett.}
  {\bf B500} (2001) 232--240,
  [\href{http://xxx.lanl.gov/abs/hep-ph/0012137}{{\tt hep-ph/0012137}}].

\bibitem{Huovinen:2001cy}
P.~Huovinen, P.~F. Kolb, U.~W. Heinz, P.~V. Ruuskanen, and S.~A. Voloshin, {\it
  {Radial and elliptic flow at RHIC: Further predictions}},  {\em Phys. Lett.}
  {\bf B503} (2001) 58--64, [\href{http://xxx.lanl.gov/abs/hep-ph/0101136}{{\tt
  hep-ph/0101136}}].

\bibitem{Kolb:2001qz}
P.~F. Kolb, U.~W. Heinz, P.~Huovinen, K.~J. Eskola, and K.~Tuominen, {\it
  Centrality dependence of multiplicity, transverse energy, and elliptic flow
  from hydrodynamics},  {\em Nucl. Phys.} {\bf A696} (2001) 197--215,
  [\href{http://xxx.lanl.gov/abs/hep-ph/0103234}{{\tt hep-ph/0103234}}].

\bibitem{Heinz:2001xi}
U.~W. Heinz and P.~F. Kolb, {\it {Early thermalization at RHIC}},  {\em Nucl.
  Phys.} {\bf A702} (2002) 269--280,
  [\href{http://xxx.lanl.gov/abs/hep-ph/0111075}{{\tt hep-ph/0111075}}].

\bibitem{Teaney:1999gr}
D.~Teaney and E.~V. Shuryak, {\it {An unusual space-time evolution for heavy
  ion collisions at high energies due to the QCD phase transition}},  {\em
  Phys. Rev. Lett.} {\bf 83} (1999) 4951--4954,
  [\href{http://xxx.lanl.gov/abs/nucl-th/9904006}{{\tt nucl-th/9904006}}].

\bibitem{Teaney:2000cw}
D.~Teaney, J.~Lauret, and E.~V. Shuryak, {\it {Flow at the SPS and RHIC as a
  quark gluon plasma signature}},  {\em Phys. Rev. Lett.} {\bf 86} (2001)
  4783--4786, [\href{http://xxx.lanl.gov/abs/nucl-th/0011058}{{\tt
  nucl-th/0011058}}].

\bibitem{Teaney:2001av}
D.~Teaney, J.~Lauret, and E.~V. Shuryak, {\it {A hydrodynamic description of
  heavy ion collisions at the SPS and RHIC}},
  \href{http://xxx.lanl.gov/abs/nucl-th/0110037}{{\tt nucl-th/0110037}}.

\bibitem{Teaney:2003kp}
D.~Teaney, {\it {Effect of shear viscosity on spectra, elliptic flow, and
  Hanbury Brown-Twiss radii}},  {\em Phys. Rev.} {\bf C68} (2003) 034913,
  [\href{http://xxx.lanl.gov/abs/nucl-th/0301099}{{\tt nucl-th/0301099}}].

\bibitem{Krasnitz:2003jw}
A.~Krasnitz, Y.~Nara, and R.~Venugopalan, {\it Classical gluodynamics of high
  energy nuclear collisions: An erratum and an update},  {\em Nucl. Phys.} {\bf
  A727} (2003) 427--436, [\href{http://xxx.lanl.gov/abs/hep-ph/0305112}{{\tt
  hep-ph/0305112}}].

\bibitem{Krasnitz:2002mn}
A.~Krasnitz, Y.~Nara, and R.~Venugopalan, {\it {Gluon production in the color
  glass condensate model of collisions of ultrarelativistic finite nuclei}},
  {\em Nucl. Phys.} {\bf A717} (2003) 268--290,
  [\href{http://xxx.lanl.gov/abs/hep-ph/0209269}{{\tt hep-ph/0209269}}].

\bibitem{Lappi:2003bi}
T.~Lappi, {\it Production of gluons in the classical field model for heavy ion
  collisions},  {\em Phys. Rev.} {\bf C67} (2003) 054903,
  [\href{http://xxx.lanl.gov/abs/hep-ph/0303076}{{\tt hep-ph/0303076}}].

\bibitem{Fukushima:2007ja}
K.~Fukushima, {\it Initial fields and instability in the classical model of the
  heavy-ion collision},  \href{http://xxx.lanl.gov/abs/0704.3625}{{\tt
  0704.3625}}.

\bibitem{Kovchegov:2007pq}
Y.~V. Kovchegov and A.~Taliotis, {\it {Early time dynamics in heavy ion
  collisions from AdS/CFT correspondence}},  {\em Phys. Rev.} {\bf C76} (2007)
  014905, [\href{http://xxx.lanl.gov/abs/0705.1234}{{\tt arXiv:0705.1234}}].

\bibitem{Kovchegov:2005ss}
Y.~V. Kovchegov, {\it {Can thermalization in heavy ion collisions be described
  by QCD diagrams?}},  {\em Nucl. Phys.} {\bf A762} (2005) 298--325,
  [\href{http://xxx.lanl.gov/abs/hep-ph/0503038}{{\tt hep-ph/0503038}}].

\bibitem{Bjorken:1982qr}
J.~D. Bjorken, {\it Highly relativistic nucleus-nucleus collisions: The central
  rapidity region},  {\em Phys. Rev.} {\bf D27} (1983) 140--151.

\bibitem{Blaizot:1987nc}
J.~P. Blaizot and A.~H. Mueller, {\it {The Early Stage of Ultrarelativistic
  Heavy Ion Collisions}},  {\em Nucl. Phys.} {\bf B289} (1987) 847.

\bibitem{McLerran:1993ni}
L.~D. McLerran and R.~Venugopalan, {\it Computing quark and gluon distribution
  functions for very large nuclei},  {\em Phys. Rev.} {\bf D49} (1994)
  2233--2241, [\href{http://xxx.lanl.gov/abs/hep-ph/9309289}{{\tt
  hep-ph/9309289}}].

\bibitem{McLerran:1993ka}
L.~D. McLerran and R.~Venugopalan, {\it Gluon distribution functions for very
  large nuclei at small transverse momentum},  {\em Phys. Rev.} {\bf D49}
  (1994) 3352--3355, [\href{http://xxx.lanl.gov/abs/hep-ph/9311205}{{\tt
  hep-ph/9311205}}].

\bibitem{McLerran:1994vd}
L.~D. McLerran and R.~Venugopalan, {\it Green's functions in the color field of
  a large nucleus},  {\em Phys. Rev.} {\bf D50} (1994) 2225--2233,
  [\href{http://xxx.lanl.gov/abs/hep-ph/9402335}{{\tt hep-ph/9402335}}].

\bibitem{Kovchegov:1996ty}
Y.~V. Kovchegov, {\it Non-abelian {Weizsaecker-Williams} field and a two-
  dimensional effective color charge density for a very large nucleus},  {\em
  Phys. Rev.} {\bf D54} (1996) 5463--5469,
  [\href{http://xxx.lanl.gov/abs/hep-ph/9605446}{{\tt hep-ph/9605446}}].

\bibitem{Kovchegov:1997pc}
Y.~V. Kovchegov, {\it Quantum structure of the non-abelian
  {Weizsaecker-Williams} field for a very large nucleus},  {\em Phys. Rev.}
  {\bf D55} (1997) 5445--5455,
  [\href{http://xxx.lanl.gov/abs/hep-ph/9701229}{{\tt hep-ph/9701229}}].

\bibitem{Kovner:1995ja}
A.~Kovner, L.~D. McLerran, and H.~Weigert, {\it Gluon production from
  non{A}belian {W}eizsacker-{W}illiams fields in nucleus-nucleus collisions},
  {\em Phys. Rev.} {\bf D52} (1995) 6231--6237,
  [\href{http://xxx.lanl.gov/abs/hep-ph/9502289}{{\tt hep-ph/9502289}}].

\bibitem{Kovchegov:1997ke}
Y.~V. Kovchegov and D.~H. Rischke, {\it Classical gluon radiation in
  ultrarelativistic nucleus nucleus collisions},  {\em Phys. Rev.} {\bf C56}
  (1997) 1084--1094, [\href{http://xxx.lanl.gov/abs/hep-ph/9704201}{{\tt
  hep-ph/9704201}}].

\bibitem{Krasnitz:1998ns}
A.~Krasnitz and R.~Venugopalan, {\it Non-perturbative computation of gluon
  mini-jet production in nuclear collisions at very high energies},  {\em Nucl.
  Phys.} {\bf B557} (1999) 237,
  [\href{http://xxx.lanl.gov/abs/hep-ph/9809433}{{\tt hep-ph/9809433}}].

\bibitem{Krasnitz:1999wc}
A.~Krasnitz and R.~Venugopalan, {\it The initial energy density of gluons
  produced in very high energy nuclear collisions},  {\em Phys. Rev. Lett.}
  {\bf 84} (2000) 4309--4312,
  [\href{http://xxx.lanl.gov/abs/hep-ph/9909203}{{\tt hep-ph/9909203}}].

\bibitem{Kovchegov:2000hz}
Y.~V. Kovchegov, {\it Classical initial conditions for ultrarelativistic heavy
  ion collisions},  {\em Nucl. Phys.} {\bf A692} (2001) 557--582,
  [\href{http://xxx.lanl.gov/abs/hep-ph/0011252}{{\tt hep-ph/0011252}}].

\bibitem{Kharzeev:2000ph}
D.~Kharzeev and M.~Nardi, {\it {Hadron production in nuclear collisions at RHIC
  and high density QCD}},  {\em Phys. Lett.} {\bf B507} (2001) 121--128,
  [\href{http://xxx.lanl.gov/abs/nucl-th/0012025}{{\tt nucl-th/0012025}}].

\bibitem{Kharzeev:2001yq}
D.~Kharzeev, E.~Levin, and M.~Nardi, {\it {The onset of classical QCD dynamics
  in relativistic heavy ion collisions}},  {\em Phys. Rev.} {\bf C71} (2005)
  054903, [\href{http://xxx.lanl.gov/abs/hep-ph/0111315}{{\tt
  hep-ph/0111315}}].

\bibitem{Kharzeev:2002pc}
D.~Kharzeev, E.~Levin, and L.~McLerran, {\it {Parton saturation and N(part)
  scaling of semi-hard processes in QCD}},  {\em Phys. Lett.} {\bf B561} (2003)
  93--101, [\href{http://xxx.lanl.gov/abs/hep-ph/0210332}{{\tt
  hep-ph/0210332}}].

\bibitem{Kharzeev:2004yx}
D.~Kharzeev, Y.~V. Kovchegov, and K.~Tuchin, {\it {Nuclear modification factor
  in d + Au collisions: Onset of suppression in the color glass condensate}},
  {\em Phys. Lett.} {\bf B599} (2004) 23--31,
  [\href{http://xxx.lanl.gov/abs/hep-ph/0405045}{{\tt hep-ph/0405045}}].

\bibitem{Albacete:2003iq}
J.~L. Albacete, N.~Armesto, A.~Kovner, C.~A. Salgado, and U.~A. Wiedemann, {\it
  {Energy dependence of the Cronin effect from non-linear QCD evolution}},
  {\em Phys. Rev. Lett.} {\bf 92} (2004) 082001,
  [\href{http://xxx.lanl.gov/abs/hep-ph/0307179}{{\tt hep-ph/0307179}}].

\bibitem{Albacete:2007sm}
J.~L. Albacete, {\it {Particle multiplicities in Lead-Lead collisions at the
  LHC from non-linear evolution with running coupling}},  {\em Phys. Rev.
  Lett.} {\bf 99} (2007) 262301, [\href{http://xxx.lanl.gov/abs/0707.2545}{{\tt
  arXiv:0707.2545}}].

\bibitem{Iancu:2003xm}
E.~Iancu and R.~Venugopalan, {\it The color glass condensate and high energy
  scattering in {QCD}},  \href{http://xxx.lanl.gov/abs/hep-ph/0303204}{{\tt
  hep-ph/0303204}}.

\bibitem{Weigert:2005us}
H.~Weigert, {\it Evolution at small {$x_{\text{bj}}$: The Color Glass
  Condensate}},  {\em Prog. Part. Nucl. Phys.} {\bf 55} (2005) 461--565,
  [\href{http://xxx.lanl.gov/abs/hep-ph/0501087}{{\tt hep-ph/0501087}}].

\bibitem{Jalilian-Marian:2005jf}
J.~Jalilian-Marian and Y.~V. Kovchegov, {\it Saturation physics and deuteron
  gold collisions at {RHIC}},  {\em Prog. Part. Nucl. Phys.} {\bf 56} (2006)
  104--231, [\href{http://xxx.lanl.gov/abs/hep-ph/0505052}{{\tt
  hep-ph/0505052}}].

\bibitem{Janik:2005zt}
R.~A. Janik and R.~Peschanski, {\it {Asymptotic perfect fluid dynamics as a
  consequence of AdS/CFT}},  {\em Phys. Rev.} {\bf D73} (2006) 045013,
  [\href{http://xxx.lanl.gov/abs/hep-th/0512162}{{\tt hep-th/0512162}}].

\bibitem{Janik:2006ft}
R.~A. Janik, {\it {Viscous plasma evolution from gravity using AdS/CFT}},  {\em
  Phys. Rev. Lett.} {\bf 98} (2007) 022302,
  [\href{http://xxx.lanl.gov/abs/hep-th/0610144}{{\tt hep-th/0610144}}].

\bibitem{Heller:2007qt}
M.~P. Heller and R.~A. Janik, {\it {Viscous hydrodynamics relaxation time from
  AdS/CFT}},  \href{http://xxx.lanl.gov/abs/hep-th/0703243}{{\tt
  hep-th/0703243}}.

\bibitem{Benincasa:2007tp}
P.~Benincasa, A.~Buchel, M.~P. Heller, and R.~A. Janik, {\it On the
  supergravity description of boost invariant conformal plasma at strong
  coupling},  \href{http://xxx.lanl.gov/abs/0712.2025}{{\tt 0712.2025}}.

\bibitem{Heller:2008mb}
M.~P. Heller, P.~Surowka, R.~Loganayagam, M.~Spalinski, and S.~E. Vazquez, {\it
  {On a consistent AdS/CFT description of boost-invariant plasma}},
  \href{http://xxx.lanl.gov/abs/0805.3774}{{\tt arXiv:0805.3774}}.
  
\bibitem{Kovner:1995ts} 
A.~Kovner, L.~D. McLerran, and H.~Weigert,
  {\it Gluon production at high transverse momentum in the
    McLerran-Venugopalan model of nuclear structure functions}, {\em
    Phys. Rev.} {\bf D52} (1995) 3809--3814,
  [\href{http://xxx.lanl.gov/abs/hep-ph/9505320}{{\tt
      hep-ph/9505320}}].

\bibitem{Landau:1953gs}
L.~D. Landau, {\it {On the multiparticle production in high-energy
  collisions}},  {\em Izv. Akad. Nauk SSSR Ser. Fiz.} {\bf 17} (1953) 51--64.

\bibitem{Bearden:2003hx}
{\bf BRAHMS} Collaboration, I.~G. Bearden {\em et.~al.}, {\it {Nuclear stopping
  in Au + Au collisions at s(NN)**(1/2) = 200-GeV}},  {\em Phys. Rev. Lett.}
  {\bf 93} (2004) 102301, [\href{http://xxx.lanl.gov/abs/nucl-ex/0312023}{{\tt
  nucl-ex/0312023}}].

\bibitem{Itakura:2003jp}
K.~Itakura, Y.~V. Kovchegov, L.~McLerran, and D.~Teaney, {\it {Baryon stopping
  and valence quark distribution at small x}},  {\em Nucl. Phys.} {\bf A730}
  (2004) 160--190, [\href{http://xxx.lanl.gov/abs/hep-ph/0305332}{{\tt
  hep-ph/0305332}}].

\bibitem{Albacete:2006vv}
J.~L. Albacete and Y.~V. Kovchegov, {\it {Baryon stopping in proton nucleus
  collisions}},  {\em Nucl. Phys.} {\bf A781} (2007) 122--149,
  [\href{http://xxx.lanl.gov/abs/hep-ph/0605053}{{\tt hep-ph/0605053}}].

\bibitem{Kovchegov:1998bi}
Y.~V. Kovchegov and A.~H. Mueller, {\it Gluon production in current nucleus and
  nucleon nucleus collisions in a quasi-classical approximation},  {\em Nucl.
  Phys.} {\bf B529} (1998) 451--479,
  [\href{http://xxx.lanl.gov/abs/hep-ph/9802440}{{\tt hep-ph/9802440}}].

\bibitem{Kopeliovich:1998nw}
B.~Z. Kopeliovich, A.~V. Tarasov, and A.~Schafer, {\it Bremsstrahlung of a
  quark propagating through a nucleus},  {\em Phys. Rev.} {\bf C59} (1999)
  1609--1619, [\href{http://xxx.lanl.gov/abs/hep-ph/9808378}{{\tt
  hep-ph/9808378}}].

\bibitem{Dumitru:2001ux}
A.~Dumitru and L.~D. McLerran, {\it How protons shatter colored glass},  {\em
  Nucl. Phys.} {\bf A700} (2002) 492--508,
  [\href{http://xxx.lanl.gov/abs/hep-ph/0105268}{{\tt hep-ph/0105268}}].

\bibitem{Kovchegov:2001sc}
Y.~V. Kovchegov and K.~Tuchin, {\it Inclusive gluon production in dis at high
  parton density},  {\em Phys. Rev.} {\bf D65} (2002) 074026,
  [\href{http://xxx.lanl.gov/abs/hep-ph/0111362}{{\tt hep-ph/0111362}}].

\bibitem{Kovchegov:2001ni}
Y.~V. Kovchegov, {\it {Diffractive gluon production in proton nucleus
  collisions and in DIS}},  {\em Phys. Rev.} {\bf D64} (2001) 114016,
  [\href{http://xxx.lanl.gov/abs/hep-ph/0107256}{{\tt hep-ph/0107256}}].

\bibitem{Brower:2007qh}
R.~C. Brower, M.~J. Strassler, and C.-I. Tan, {\it {On the Eikonal
  Approximation in AdS Space}},  \href{http://xxx.lanl.gov/abs/0707.2408}{{\tt
  arXiv:0707.2408}}.

\bibitem{Levin:2008vj}
E.~Levin, J.~Miller, B.~Z. Kopeliovich, and I.~Schmidt, {\it {Glauber - Gribov
  approach for DIS on nuclei in N=4 SYM}},
  \href{http://xxx.lanl.gov/abs/0811.3586}{{\tt arXiv:0811.3586}}.

\bibitem{Cornalba:2007zb}
L.~Cornalba, M.~S. Costa, and J.~Penedones, {\it {Eikonal Approximation in
  AdS/CFT: Resumming the Gravitational Loop Expansion}},  {\em JHEP} {\bf 09}
  (2007) 037, [\href{http://xxx.lanl.gov/abs/0707.0120}{{\tt
  arXiv:0707.0120}}].

\bibitem{Cornalba:2006xm}
L.~Cornalba, M.~S. Costa, J.~Penedones, and R.~Schiappa, {\it {Eikonal
  approximation in AdS/CFT: Conformal partial waves and finite N four-point
  functions}},  {\em Nucl. Phys.} {\bf B767} (2007) 327--351,
  [\href{http://xxx.lanl.gov/abs/hep-th/0611123}{{\tt hep-th/0611123}}].

\bibitem{Cornalba:2006xk}
L.~Cornalba, M.~S. Costa, J.~Penedones, and R.~Schiappa, {\it {Eikonal
  approximation in AdS/CFT: From shock waves to four-point functions}},  {\em
  JHEP} {\bf 08} (2007) 019,
  [\href{http://xxx.lanl.gov/abs/hep-th/0611122}{{\tt hep-th/0611122}}].

\bibitem{deHaro:2000xn}
S.~de~Haro, S.~N. Solodukhin, and K.~Skenderis, {\it {Holographic
  reconstruction of spacetime and renormalization in the AdS/CFT
  correspondence}},  {\em Commun. Math. Phys.} {\bf 217} (2001) 595--622,
  [\href{http://xxx.lanl.gov/abs/hep-th/0002230}{{\tt hep-th/0002230}}].

\bibitem{Krasnitz:2003nv}
A.~Krasnitz, Y.~Nara, and R.~Venugopalan, {\it Probing a color glass condensate
  in high energy heavy ion collisions},  {\em Braz. J. Phys.} {\bf 33} (2003)
  223--230.

\bibitem{Albacete:2008ze}
J.~L. Albacete, Y.~V. Kovchegov, and A.~Taliotis, {\it {DIS on a Large Nucleus
  in AdS/CFT}},  {\em JHEP} {\bf 07} (2008) 074,
  [\href{http://xxx.lanl.gov/abs/0806.1484}{{\tt arXiv:0806.1484}}].

\bibitem{Danielsson:1998wt}
U.~H. Danielsson, E.~Keski-Vakkuri, and M.~Kruczenski, {\it {Vacua,
  Propagators, and Holographic Probes in AdS/CFT}},  {\em JHEP} {\bf 01} (1999)
  002, [\href{http://xxx.lanl.gov/abs/hep-th/9812007}{{\tt hep-th/9812007}}].

\bibitem{Chesler:2008hg}
P.~M. Chesler and L.~G. Yaffe, {\it {Horizon formation and far-from-equilibrium
  isotropization in supersymmetric Yang-Mills plasma}},
  \href{http://xxx.lanl.gov/abs/0812.2053}{{\tt arXiv:0812.2053}}.

\end{thebibliography}

\providecommand{\href}[2]{#2}\begingroup\raggedright\endgroup


\end{document}